\documentclass{thesisclass}
\pdfoutput=1

\hypersetup{
 pdfauthor={Michael Tänzer},
 pdftitle={The Influence of Architectural Styles on Security, Using the Example of a Certification Authority},
 pdfsubject={The Influence of Architectural Styles on Security, Using the Example of a Certification Authority},
 pdfkeywords={Software Architecture, Architectural Style, Security},
 pdfborder={0 0 0}
}

\newcommand{\myname}{Michael Tänzer}
\newcommand{\mytitle}{The Influence of Architectural Styles on Security, Using the Example of a Certification Authority}
\newcommand{\myinstitute}{Institute for Program Structures\\
													and Data Organization (IPD)}

\newcommand{\reviewerone}{Prof. Dr. Ralf H. Reussner}
\newcommand{\reviewertwo}{Prof. Dr. Walter F. Tichy}
\newcommand{\adviser}{M.~Sc. Zoya Durdik}
\newcommand{\advisertwo}{Dipl.-Inform. Matthias Huber}

\newcommand{\timestart}{15. May 2013}
\newcommand{\timeend}{17. July 2013}
\newcommand{\submissiontime}{16. July 2013}

\newcommand{\todo}[1]{}

\usepackage[toc]{glossaries}
\newcommand{\myacronym}[3]{\newacronym[description={(#2) #3}]{#1}{#1}{#2}}

\newglossaryentry{AStyle}
{
	name={architectural style},
	description={(sometimes also called ``architectural pattern'') A common design that is applied to the whole architecture of a system. It often specifies how components of the system interact with each other.\\
	Examples of architectural styles: Pipes and Filters, Client/Server, Layered Architecture, Service-Oriented Architecture},
}

\newglossaryentry{Assurance}
{
	name={Assurance},
	description={The process by which the \gls{Assurer} verifies the identity of an \gls{Assuree}. Sometimes it also refers to the data record created in the web application for an Assurance},
}

\newglossaryentry{Assurance points}
{
	name={Assurance points},
	description={For each verification of the identity of an \gls{Assuree} the \gls{Assurer} assigns a confidence value, which is expressed in Assurance points. The amount of Assurance points the \gls{Assurer} may issue depends on his experience},
}

\newglossaryentry{Assuree}
{
	name={Assuree},
	description={CAcert community member that wants to get her identity verified},
}

\newglossaryentry{Assurer}
{
	name={Assurer},
	description={CAcert community member that is already well-verified. He will check the documents of \glspl{Assuree} to verify their identity},
}

\newacronym{CA}{CA}{certification authority}

\myacronym{CERT}{computer emergency response team}
{often refers to the CERT Coordination Center located at the Carnegie Mellon University, which issues advisories to system administrators and developers about common existing or potential threats to computer security}

\myacronym{CSRF}{cross-site request forgery}
{is an attack where malicious data is placed on a third party website that causes a user agent displaying that third party website to make a cross-site request to the target website possibly without the user knowing about the request being made. This forged request includes authentication state such as cookies if the user has been previously logged in to the website and may lead to side effects which seem to happen on behalf of the user such as a transaction being made, an item being purchased, a message being sent etc}

\myacronym{CVE}{common vulnerabilities and exposures}
{is an index of publicly known information-security vulenarabilities and exposures}

\myacronym{TLS}{transport layer security}{specified in \cite{rfc5246} is a successor to the SSL (Secure Sockets Layer) protocol. It provides an authenticated, confidential connection oriented transport service for application layer protocols. It is best known for its use with HTTP to securely deliver web sites}

\newglossaryentry{x509}
{
	name={X.509},
	description={A standard format for certificates \cite{ITU2011}. It contains identity information about the subject identified and its public key and can be extended in various ways. Many services use X.509 certificates to establish the identity of the other party involved in communication, most noteworthy \gls{TLS} and S/MIME (a format for securing email communication)},
}

\makeglossaries

\usepackage{amsthm}
\theoremstyle{definition}
\newtheorem{defn}{Definition}

\usepackage{enumitem}
\newlist{ind_legal}{enumerate}{10}
\setlist[ind_legal]{label*=\arabic*.}

\newlength\labwd
\makeatletter

\newcounter{enumv}[enumiv]
\newcounter{enumvi}[enumv]
\newcounter{enumvii}[enumvi]
\newenvironment{legal}
	{\advance\@enumdepth\@ne
	\ifnum \@enumdepth >7\@toodeep\else
	\edef\@enumctr{enum\romannumeral\the\@enumdepth}

	\renewcommand{\p@enumi}{}

	\renewcommand{\p@enumii}{}

	\renewcommand{\p@enumiii}{}

	\renewcommand{\p@enumiv}{}

	\renewcommand{\p@enumv}{}

	\renewcommand{\p@enumvi}{}

	\renewcommand{\p@enumvii}{}
	\begin{list}{\csname label\@enumctr\endcsname}{%
		\usecounter\@enumctr
		\setlength\labelwidth{\z@}
		\setlength\labelsep{\z@}
		\setlength\leftmargin{15pt}
		\setlength\labwd{\ifcase \@enumdepth \or -15pt\or -30pt\or -45pt\or -60pt\or -75pt\or -90pt\or -105pt\fi}
		
	}%
	\fi
	}
	{\ifnum \@enumdepth >7\else\end{list}\fi}
\makeatother

\usepackage{xspace}
\newcommand*{\eg}{e.g.\@\xspace}
\newcommand*{\ie}{i.e.\@\xspace}
\makeatletter
\newcommand*{\etc}{%
	\@ifnextchar{.}%
		{etc}%
		{etc.\@\xspace}%
}
\makeatother

\usepackage{marginnote}

\newcommand{\newreq}
{\marginnote{New\\Feature}}
\newcommand{\oldreq}
{\marginnote{Existing\\Feature}}
\newcommand{\altreq}
{\marginnote{Altered\\Feature}}

\newcommand{\riskheader}
{\marginnote{\textbf{Effort/Risk/\\Gain}}}
\newcommand{\risk}[3]
{\marginnote{\textbf{#1/#2/#3}}}

\begin{document}

\selectlanguage{english}

\frontmatter
\pagenumbering{roman}

\newcommand{\diameter}{20}
\newcommand{\xone}{-15}
\newcommand{\xtwo}{160}
\newcommand{\yone}{15}
\newcommand{\ytwo}{-253}

\begin{titlepage}
\begin{tikzpicture}[overlay]
\draw[color=gray]  
 		 (\xone mm, \yone mm)
  -- (\xtwo mm, \yone mm)
 arc (90:0:\diameter pt) 
  -- (\xtwo mm + \diameter pt , \ytwo mm) 
	-- (\xone mm + \diameter pt , \ytwo mm)
 arc (270:180:\diameter pt)
	-- (\xone mm, \yone mm);
\end{tikzpicture}
	\begin{textblock}{10}[0,0](4,2.5)
		\includegraphics[width=.3\textwidth]{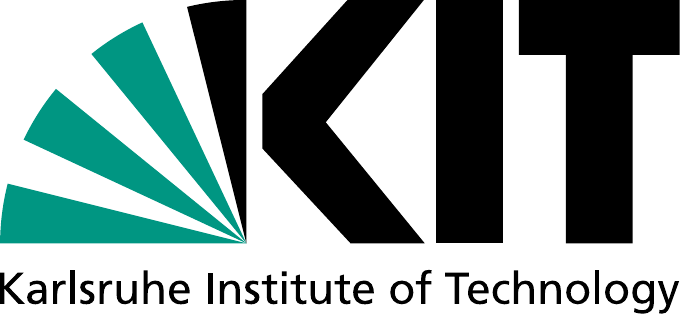}
	\end{textblock}
	\changefont{phv}{m}{n}	
	\vspace*{3.5cm}
	\begin{center}
		\Huge{\mytitle}
		\vspace*{2cm}\\
		\Large{
			\iflanguage{english}{Study Thesis by}
												{Studienarbeit von}
		}\\
		\vspace*{1cm}
		\huge{\myname}\\
		\vspace*{1cm}
		\Large{
			\iflanguage{english}{At the Department of Informatics}			
													{An der Fakult\"at f\"ur Informatik}
			\\
			\myinstitute
		}
	\end{center}
	\vspace*{1cm}
\Large{
\begin{center}
\begin{tabular}[ht]{l c l}
  \iflanguage{english}{Reviewer}{Erstgutachter}: & \hfill  & \reviewerone\\
  \iflanguage{english}{Second reviewer}{Zweitgutachter}: & \hfill  & \reviewertwo\\
  \iflanguage{english}{Adviser}{Betreuender Mitarbeiter}: & \hfill  & \adviser\\
  \iflanguage{english}{Second adviser}{Zweiter betreuender Mitarbeiter}: & \hfill  & \advisertwo\\
\end{tabular}
\end{center}
}

\vspace{2cm}
\begin{center}
\large{\iflanguage{english}{Duration}{Bearbeitungszeit}: \timestart \hspace*{0.25cm} -- \hspace*{0.25cm} \timeend}
\end{center}

\begin{textblock}{10}[0,0](4,16.8)
\tiny{ 
	\iflanguage{english}
		{KIT -- University of the State of Baden-Wuerttemberg and National Research Center of the Helmholtz Association}
		{KIT -- Universit\"at des Landes Baden-W\"urttemberg und nationales Forschungszentrum in der Helmholtz-Gemeinschaft}
}
\end{textblock}

\begin{textblock}{10}[0,0](14,16.75)
\large{
	\textbf{www.kit.edu} 
}
\end{textblock}

\end{titlepage}

\vspace*{36\baselineskip}
\hbox to \textwidth{\hrulefill}
\par
\iflanguage{english}{I declare that I have developed and written the enclosed thesis completely by myself, and have not used sources or means without declaration in the text.}{Ich versichere wahrheitsgem\"a\ss, die Arbeit selbstst\"andig angefertigt, alle benutzten Hilfsmittel vollst\"andig und genau angegeben und alles kenntlich gemacht zu haben, was aus Arbeiten anderer unver\"andert oder mit Ab\"anderungen entnommen wurde.}

\textbf{Karlsruhe, \submissiontime}
\vspace{1.5cm}

\dotfill\hspace*{8.0cm}\\
\hspace*{2cm}(\textbf{Michael Tänzer}) 

\thispagestyle{empty}

\blankpage


\chapter{Abstract}

Often, security is considered in an advanced stage of the implementation of a system, rather than integrating it into the system design. This leads to less secure systems, as the security mechanisms are only applied as an afterthought and therefore do not integrate well with the rest of the design. Also, several statistics about discovered vulnerabilities in existing systems suggest, that most of the vulnerabilities of a system are not caused by errors in the cryptographic primitives, but in other parts of the implementation. So integrating security concerns early in the design process seems a promising approach for increasing the security of the resulting system.

This work evaluates how the choice of the \gls{AStyle} affects the security of the resulting system. The evaluation is done on the example of an existing \gls{CA}. The requirements for the system are gathered and multiple designs according to different \glspl{AStyle} are drafted and evaluated using a risk evaluation method. Then the evaluated designs are compared to find out whether there are significant differences.

\chapter{Zusammenfassung}

Häufig wird die Sicherheit eines Systems erst in einem fortgeschrittenen Entwicklungs\-stadium berücksichtigt, statt dies in den Entwurfsprozess zu integrieren. Dies führt zu unsichereren Systemen, da die Sicherheitsmechanismen erst im Nachhinein hinzugefügt werden und deshalb nicht gut in den bestehenden Systementwurf integriert sind. Außerdem lassen einige Statistiken über entdeckte Sicherheitslücken in existierenden Systemen vermuten, dass die meisten Fehler die zu Sicherheitslücken führen, nicht in den kryptographischen Primitiven zu finden sind, sondern in anderen Teilen der Implementierung. Deshalb scheint die Integration von Sicherheits-Aspekten früh im Entwurfsprozess vielversprechend um die Sicherheit des resultierenden Systems zu verbessern.

Diese Arbeit evaluiert, wie sich die Entscheidung für einen Architekturstil auf die Sicherheit des resultierenden Systems auswirkt. Die Bewertung wird am Beispiel einer existierenden Zertifizierungsstelle (CA) ausgeführt. Die Anforderungen an das System werden erhoben und es werden mehrere Entwürfe anhand verschiedener Architekturstile angefertigt und anhand einer Risiko-Bewertungsmethode evaluiert. Anschließend werden die evaluierten Entwürfe verglichen um festzustellen ob es signifikante Unterschiede gibt.

\tableofcontents
\blankpage

\mainmatter
\pagenumbering{arabic}

\chapter{Introduction}
\label{ch:Introduction}

The lack of design with security in mind has often been discussed \cite[p.~38]{fernandez2001,anderson2001,stallings2011}. As a solution, processes that take security into account in all aspects of the software development process and try to avoid this ``security as an afterthought'' problem have been proposed \cite{mouratidis2003}. Additionally, statistics of vulnerabilities found in existing software systems suggest that it is not so much deficiencies in using cryptographic primitives, but errors in the general implementation of the software that cause most of the vulnerabilities. According to \cite[p.~110]{schneider1999} less than 15\% of all \gls{CERT} advisories could have been avoided by proper use of cryptography and statistics of the \gls{CVE} index \cite{christey2007} indicate that only 1.5\% of all vulnerabilities were caused by errors in parts of the software concerning cryptography. Those vulnerabilities may also partly stem from a design that has the tendency to be less secure when implemented.

These observations led to the consideration of evaluating different \glspl{AStyle} with respect to the security likely to be achieved by an implementation of these styles. Because such an evaluation is pretty hard if not impossible to do on a ``general idea'' of the \gls{AStyle}, a real system with real requirements (some of the main non-functional requirements relating to security of the system) is chosen and some designs according to different \glspl{AStyle} are crafted and evaluated. The concrete system selected for this work is a \gls{CA} based on a web of trust, named CAcert, which is described in Section \ref{sec:CAcert}. CAcert was chosen because of the high security requirements needed in this environment and the author's familiarity with that system.

This work focuses on the design on the architectural level and evaluating the resulting design, it does not go further into a detailed design or even an implementation stage. Due to this limitation of scope a formal validation of the approach is not conducted. Also due to time constraints only two \glspl{AStyle} were evaluated, however this is not an inherent limit and the method provided allows to compare more styles with each other.

This work is structured as follows. Chapter~\ref{ch:Foundations} introduces basic concepts and gives an overview of related work. Chapter~\ref{ch:Approach} describes the approach used in this work. In Chapter~\ref{ch:Alternatives}, several designs according to different architectural styles are drafted, which are then evaluated in Chapter~\ref{ch:Evaluation}. Finally in Chapter~\ref{ch:Conclusion}, the conclusions that can be drawn from the evaluation, are presented.


\chapter{Foundations}
\label{ch:Foundations}

This work is based on various concepts from other works which are introduced in this chapter. In Section \ref{sec:Definitions} definitions about what constitutes an architectural style and basic security properties are established. Section \ref{sec:CAcert} describes the CAcert system in more detail, from the basic principle of operation to a description of the existing software and its shortcomings. The attack tree method used for evaluating the designs is presented in Section \ref{sec:EvalMethod}, and Section \ref{sec:Literature} lists work related to the topic of this paper.

\section{Definitions}
\label{sec:Definitions}

\begin{defn}[Architectural Style]
	``If architecture is a formal arrangement of architectural elements, then architectural style is that which abstracts elements and formal aspects from various specific architectures. An architectural style is less constrained and less complete than a specific architecture. [\dots] The important thing about an architectural style is that it encapsulates important decisions about the architectural elements and emphasizes important constraints on the elements and their relationships. The useful thing about style is that we can use it both to constrain the architecture and to coordinate cooperating architects.'' \cite{perry1992}
\end{defn}


\begin{defn}[Confidentiality]
	``The property that information is not disclosed to system entities (users, processes, devices) unless they have been authorized to access the information.'' \cite{NIAG2010}
\end{defn}

\begin{defn}[Integrity]
	``The property whereby an entity has not been modified in an unauthorized manner.'' \cite{NIAG2010}
\end{defn}

\begin{defn}[Availability]
	``The property of being accessible and useable upon demand by an authorized entity.'' \cite{NIAG2010}
\end{defn}

\section{CAcert}
\label{sec:CAcert}

CAcert is a free\footnote{Both free as in freedom and as in free of costs} \gls{CA}, driven by an open community. That means it offers \gls{x509} and other certificates that users can use for various purposes including securing network traffic with the \gls{TLS} protocol.

CAcert operates on a non-commercial basis which is made possible by a web of trust run by the community. Prior to getting a certificate issued under her name, a user has to get her identity verified. To achieve this, she meets an already well-verified community member, an \gls{Assurer}. The Assurer verifies her identity documents (this process is called \gls{Assurance}) and records the fact that he has done so on paper (which he archives) and in the web application. With each \gls{Assurance} the \gls{Assuree} gets a number of Assurance Points, based on the documents provided and the experience of the \gls{Assurer}. When the user has got enough points (which requires at least two \glspl{Assurance}) she can get certificates issued on her name. After further verification (at least three \glspl{Assurance} in total) and a short test on the knowledge about the \gls{Assurance} process she may become an \gls{Assurer} herself and verify other users. The more \glspl{Assurance} an \gls{Assurer} has already done, the higher his experience is assumed to be and therefore the more \gls{Assurance} points he may issue.

When there are disputes or irregularities in the community, such as mistakes made during an \gls{Assurance}, they are brought into arbitration. In arbitration senior members of the community, called Arbitrators, review the case, taking into account existing CAcert policies and principles, and give a ruling which is binding to all members.

Most of the core functionality of CAcert is handled by a custom software system, which is described in the following sections.

\subsection{Features of the Existing System}
\label{sec:CAcert:sec:Existing}

Some of the main features of the existing software are the following:
\begin{itemize}
	\item Common functionality not tied to a user account
	\begin{itemize}
		\item Informational web pages\\
			Display the home page, help pages, \etc.
		\item Contact support staff\\
			Send a message to the support team, who can help with account issues and other problems.
	\end{itemize}
	
	\item User account management
	\begin{itemize}
		\item Register new account
		\item Change account details
		\item Reset password\\
			Allows to reset the password if it has been forgotten, using lost password questions and an email confirmation for authentication.
		\item Organisational accounts\\
			For companies and other officially registered organisations, with a possibility to have more than one administrator managing the account and including the organisation name in the certificates.
	\end{itemize}
	
	\item Web of trust
	\begin{itemize}
		\item Enter \glspl{Assurance}\\
			Allows an \gls{Assurer} to check the information present in the system against the information gathered in the \gls{Assurance} and allocate Assurance points.
		\item Check verification level of a user\\
			Shows how many Assurance and experience points a user has gathered and how this value is derived.
		\item Find \glspl{Assurer} in the \gls{Assuree}'s area\\
			Helps a new user to find \glspl{Assurer} in her area and contact them in order to meet and get assured.
	\end{itemize}
	
	\item Certificate issuing
	\begin{itemize}
		\item Verify alternative identities (email addresses, domains etc.)\\
			Allows to manage email addresses and domain names, which will be verified when added to prevent abuse (by sending a verification email to the specified address).
		\item Accept certificate requests and generate certificates\\
			Allows users to submit a certificate request for a previously verified identity and to retrieve the resulting certificate.
		\item Remind users of expiring certificates\\
			Users are reminded to renew their certificates before they expire (one of the most common problems on sites using encrypted connections).
		\item Revoke certificates\\
			If a key might have been exposed or if it was lost, users should revoke the corresponding certificate so that other parties get notified of the problem when they are about to use the certificate.
		\item Renew certificates\\
			Issue a new certificate for the same key pair and with the same details as in the existing certificate but with an updated validity period.
	\end{itemize}
	
	\item Administrative functions
	\begin{itemize}
		\item Access user details\\
			Search for an account and view its details.
		\item Password reset by the support staff
		\item Change some user details\\
			Allows the support staff to change account data like the name and day of birth even after an \gls{Assurance} already has been issued, which for example might be needed on request of an Arbitrator.
		\item View and revoke Assurances\\
			Support staff may have to revoke an \gls{Assurance} on request of an Arbitrator or if an Assurer realises he made a mistake just after entering the \gls{Assurance}.
		\item Lock down accounts\\
			If an account is to be terminated or subject to further investigation in an arbitration case, access to it can be blocked.
	\end{itemize}
\end{itemize}

\subsection{Problems of the Existing System}
\label{sec:CAcert:sec:Problems}

The existing software already implements the basic features but maintaining and reviewing is difficult because the design is of poor quality. The structure loosely resembles some kind of Model View Controller architecture but the components are not well-separated and the model only consists of a plain SQL database with no abstraction on top of it (model classes, object relational mapping or something in that direction). As a consequence of the missing model abstraction, the controllers are huge and even the views have some logic in them apart from presentational concerns. The same pieces of logic (often containing raw SQL statements) appear in many places of the software, so changing behaviour requires quite a few changes.

The software is not well-modularized. The functionality that does not require an user account and the web of trust functionality are separated. The rest, however, is located in one big controller which is over 3000 lines of PHP code, largely without any comments. Even in these large files, there is almost no use of functions or methods and therefore almost no code reuse. As a consequence, the code resembles what commonly is referred to as ``spaghetti code''. The software doesn't use object-oriented programming paradigms and no framework or abstraction library for web application development is used apart from the ones built into PHP. There are only a few libraries used at all and as strings are not encoded in Unicode, proper internationalisation is a major problem.

Another issue for the whole CAcert project is that the CAcert root certificate is not yet included in the major browsers by default. So users who don't know about CAcert or haven't imported the CAcert root certificate into their browser will get a warning message when visiting a website secured with a CAcert issued certificate. In order to solve that problem CAcert has to get audited. That means that an independent Auditor reviews all policies and other documents and checks if the procedures actually implemented in CAcert (both in software as well as manual processes) fit the requirements set by the policies. This Audit also contains a code review of the critical parts of the system. This code review, however, would probably take a lot of effort because of the code quality of the existing implementation, as already described.

All in all the software is in such a bad state that a complete redesign seems unavoidable in order to review and maintain it.

A software workshop held by CAcert developers in April 2009 came to a similar conclusion. Some ideas have been exchanged, some requirements and best practises named and an idea for a design approach has been proposed. This idea however was so rough that it didn't provide enough guidance for the implementation phase that should have followed. There were some efforts to get the implementation going but they died soon after they had started.

\section{Method for Evaluation}
\label{sec:EvalMethod}

For the evaluation of the designs, the attack tree analysis is used (according to \cite{viega2001}, based on the fault tree analysis described by \cite{leveson1995}). It tries to cover all known attacks possible on the system in a systematic fashion: by grouping them into a tree structure and then assign a probability to them being successful. In this tree structure the root represents the main goal the attacker is trying to achieve. Each child node defines a more detailed possibility to achieve the goal of its parent, eventually resulting in leaf nodes representing concrete attacks. Usually child nodes are to be read as options meaning that attaining any goal of a child node results in the parent goal being achieved (logical OR), but there can also be nodes where all goals of the child nodes have to be attained in order to achieve the parent goal – this is marked by the phrase AND on every but the last child node. When the tree has been completed, a risk value is assigned beginning from the leaves. The risk values of each child node can be cumulated into a risk value for the parent node. For OR nodes this is most likely the risk value of the child with the highest risk while for AND nodes it is the lowest. However if there are many OR child nodes it may be more realistic to assign a higher risk value to the parent to account for the many possibilities to achieve the goal where only one successful attack is needed from an attacker's perspective and vice versa for many AND child nodes.

According to  Viega and McGraw the attack tree analysis has the drawback that it is subjective to a certain degree \cite{viega2001}. Firstly, the completeness of the attack tree depends on the knowledge about possible attacks, yet unknown attacks are not accounted for, and secondly assigning the risk values is mainly an educated guess where in some cases evidence may be gathered to support it. So in conclusion the accuracy of the attack tree analysis is better when carried out by a team of experts.

The problem is, that measuring the security of a system, especially during the development or even design phase, is difficult. Even more so because objective metrics that are widely-known, accurate, and proven, do not exist in that area. So the attack tree analysis was chosen for its practicality.

\newpage 
\section{Related Work}
\label{sec:Literature}

Mouratidis et al.\ \cite{mouratidis2003} try to approach security with a method derived from Tropos covering the whole requirement analysis, software design, and development process. They model the system as a set of related actors, secure entities (goals, tasks and resources), secure dependencies and security constraints. Although their main contribution is to introduce a process that uses the same concepts and notations for the entire development process, the models used in each development stage are very different and can not be easily obtained just by transforming and enriching the diagram from the previous stage. Therefore, the benefit of using the same notation in each development stage can be seen as less significant and choosing a specialised representation for each stage might have advantages. Furthermore most of their proposed process is represented in diagrams, which can get quite complicated, and they have not evaluated their approach.

Starting with Yoder and Barcalow \cite{yoder1998}, various security patterns have been proposed to equip software developers with reusable building blocks to solve security problems \cite{schumacher2001,fernandez2001,schumacher2006}. But they mostly address a single specific problem and thus can be seen as building blocks, not so much as an approach for designing a whole system.

Fielding \cite{fielding2000} provides an overview over various architectural styles often used in network-based software, with regard to performance, scalability, simplicity, modifiability, visibility, portability and reliability, but he did not explicitly cover security.

Viega and McGraw \cite{viega2001} introduce some basic knowledge about developing secure software and risk assessment, including the attack tree analysis used in this work. They stress the importance of doing a security analysis in the design phase to find flaws on the architectural level, which are hard to fix later on without introducing new flaws, but they do not explicitly cover \glspl{AStyle}.

In general, there are many approaches that propose solutions to recurring problems, as in the pattern community, or even a whole design process, but a systematic or even formal evaluation is often missing. Also the expectation, that there will be errors in the implementation, and one of the goals of a good design with security in mind is to confine the impact of these errors, is often not reflected in the description of the patterns and processes.

\chapter{Approach}
\label{ch:Approach}

This work is structured in three phases, as depicted in Figure \ref{fig:approach}: First, there is a detailed requirement analysis, which apart from the functional requirements also contains non-functional requirements (including security) and a risk assessment. The second phase is an iterative evaluation of various \glspl{AStyle}, followed by the third phase in which the styles from the previous phase are compared, to find out whether the choice of the \gls{AStyle} bears any significance and which one is the most suitable for the example system. A detailed description of each phase is given in the following sections.

\begin{center}
	\setlength\fboxsep{0pt}
	\setlength\fboxrule{0.5pt}
		\includegraphics[width=1.00\textwidth,trim=1.1cm 9.4cm 1.2cm 1.1cm,clip]{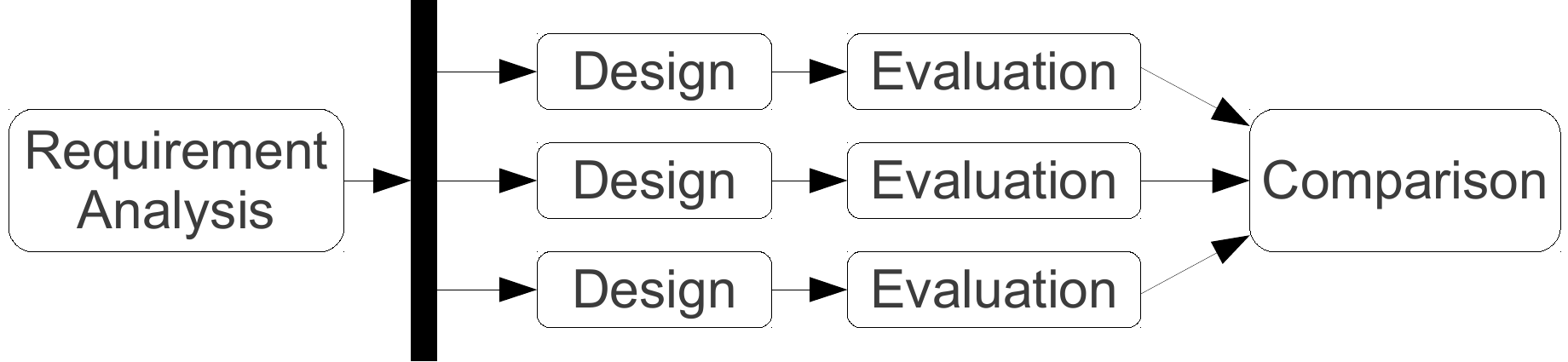}
	\captionof{figure}{Schematic Illustration of the Approach}
	\label{fig:approach}
\end{center}

\section{Requirement Analysis}
\label{sec:Requirements}

There was no existing requirements document describing all functional and non-functional requirements for the system. As a thorough understanding of the requirements provides the base for the design, a requirement analysis was conducted.

The requirements were derived from various sources: Most of the functionality of the existing system should be preserved except where good arguments are provided why the functionality is disadvantageous, unnecessary, or should be implemented outside the scope of the system. Additionally, there exist quite a few policies from CAcert or affecting CAcert that also have implications for the software. These should be respected, although policies are not immutable and if there is a good reason raised in the design process, they may be changed (at least the CAcert-driven ones). Another source of requirements were various ideas about new features that could not easily be implemented in the existing system. Some of those ideas were documented in the previous design attempt mentioned in Section \ref{sec:CAcert:sec:Existing}, others were mentioned in bug reports or posts on mailing lists discussing CAcert. Finally there are also many domain experts in the community (including the author of this thesis) who were asked for additional requirements not yet covered by the other sources.

In addition to the current functional and non-functional requirements for the system, the requirement analysis also tried to cover future requirements, \ie requirements that are currently not present but where it is probable that a change of requirements may be needed in the future.

Adding directly to the requirement analysis a risk assessment was performed (roughly following \cite{schumacher2006}). First, the main critical assets were identified. Then, the needed kind of protection for each asset (confidentiality, integrity etc.) as well as the estimated damage if this protection fails were identified. The result is a document which can be found in Appendix~\ref{Requirements} that also codifies some security requirements or can be used to extract them.

The resulting requirements document is rather detailed to facilitate the design process that followed. The highly detailed specification of the requirements also helped to discover discrepancies in the requirements themselves and probably avoided problems in subsequent development stages.

\section{Evaluation of Architectural Styles}
\label{sec:Evaluation}

Following the requirement analysis, several \glspl{AStyle} were evaluated in order to be able to compare them. The process for evaluation was as follows.

First, a style was selected for evaluation and a rough assessment just on the known properties of the style was done, for example in regard to whether it is suitable to fit the functional and non-functional requirements or whether it has unacceptable implications. Afterwards a rough design following the style (if it wasn't already classified as unsuitable) was produced, taking into account the functional and non-functional requirements and incorporating safeguards against assumed security risks. The resulting designs can be found in Chapter \ref{ch:Alternatives}.

Then the designs were evaluated based on the attack tree method and ``common sense''. The results of the evaluation and a more thorough description of the evaluation process is provided in Chapter \ref{ch:Evaluation}.

\section{Comparison of the Architectural Styles}
\label{sec:Comparison}

The evaluations from the previous phase were studied to find differences in the risk values. Where differences were found it was tried to retrace where they stemmed from, for example if there were any structural differences, implied by properties of the \gls{AStyle}, that caused the differences in the risk evaluation. This required additional work on the existing designs of the compared styles if the rough nature of the designs didn't allow for a detailed enough analysis. Based on this comparison it was determined if the choice of the \gls{AStyle} had a significant impact on the attack risk of the resulting design.

In addition to this comparison, a decision was made which design  was the most suitable for the example system. This decision did not only take security requirements but all functional and non-functional requirements into account.


\chapter{Architectural Design Alternatives}
\label{ch:Alternatives}

This chapter describes the \glspl{AStyle} selected for evaluation and the design that was created for each style. For each style that was considered a description of the style is given and it is explained why the \glspl{AStyle} layered architecture and service-oriented architecture were chosen for evaluation and why the pipes and filters architecture was not pursued further. Also a detailed description of the resulting design according to the two selected \glspl{AStyle} is given.

\section{Layered Architecture}

The first \gls{AStyle} selected to be evaluated was the layered architecture style. A layered architecture is characterised by the separation of components into different levels of abstraction, called layers. The upper layers use the lower layers to provide functionality, but never the other way around. When using a strict layered architecture a component may only use a component in the layer directly beneath it while in a relaxed layered architecture a component may use components in any lower layer (\ie skip a layer). Layering may be enforced by using runtime protection mechanisms such as the CPU privilege levels (utilised in protection of operating systems) or separating the layers in tiers, which means that the layers are grouped and each group is executed in a separate process or even on different machines, which interact in a client-server style manner, with the lower layer being the server for the upper layer. A more detailed description of the layered architecture pattern can be found in \cite{buschmann1996}.

\subsection{Choice}
\label{eval_layer_choice}

The layered architecture pattern was selected because it is often used in contexts where security plays a major role. For example operating systems are often structured in layers, separating the user space from the kernel space and possibly have further layers within the operating system kernel. Also quite a few typical business applications are designed as a multi-tier architecture.

\subsection{Design}
\label{eval_layer_design}

\begin{figure}[ht]
	\centering
	\includegraphics[width=1.00\textwidth, trim=0.8cm 3.4cm 2.5cm 0.5cm]{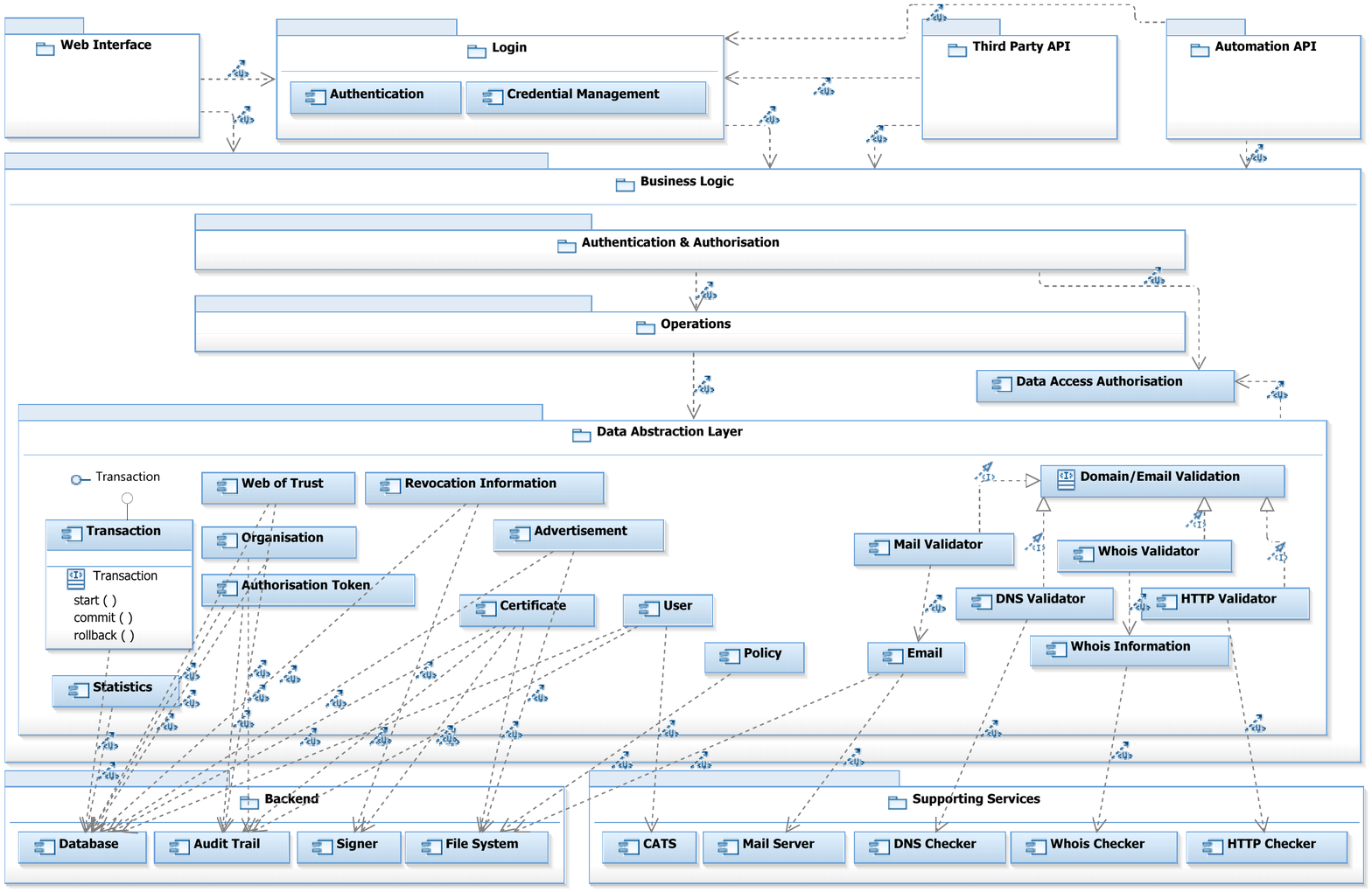}
	\caption{Layered architecture design – larger version in Appendix \ref{layer_design}}
	\label{fig:layers}
\end{figure}

The system is divided into five layers using strict layering (\ie no layer may be skipped). Strict layering was used in order to confine the effects of an attack against upper layers. So an exploit of one of the upper layers can not be used to directly access and attack the lowest layer. The resulting architecture design can be found in Figure \ref{fig:layers}. The following sections describe the different layers in more detail.

\subsubsection{Front-End Layer}
\label{frontend_layer}
The topmost layer is the front-end layer and consists of all components directly facing the external network. These components are further divided vertically into the web interface, the login server, and remote APIs for third parties. The authentication and credential management components were separated from the rest to isolate them, assuming that the code base may be rather small and better reviewed, resulting in a smaller attack surface on the login credentials. Other components will not directly work on the credentials and only get an authentication token from the client, which they can use to verify that the client has been successfully authenticated by the authentication component by querying the authentication and authorisation layer. Other front-end components still have to check the re-authentication credentials though. The front-end components realise all functionality they can not handle by themselves by making a request to the lower layer, forwarding the authentication token for proof of authentication.

The front-end components do not keep any state so they can be scaled up by deploying components multiple times, without needing to migrate state between machines, and it gets more difficult to expose restricted information from another user by attacking the front-end components. If there is interaction context that can not be held by the client (by using cookies or request parameters), which for example might be the case with \gls{CSRF} protection mechanisms, it should be forwarded for to lower layers for storage.

\subsubsection{Authentication and Authorisation Layer}
\label{auth_layer}
The second layer is the authentication and authorisation layer. The authentication state of a client is forwarded to this layer by the authentication component in the front-end layer on successful authentication of the client, including all information needed to validate the authentication token provided by the client on successive requests. For each request coming from the front-end layer, the validity of the authentication token and whether the authenticated user is authorised to execute operation requested is checked. If the authorisation is granted, the request is forwarded to the operations layer, including information about the authentication state.

The authentication and authorisation layer should only keep soft state, meaning state that is recoverable and that is discarded after some time-out or on demand. This allows some possibly expensive to reproduce state to be cached and still makes it possible to deploy the layer on multiple machines. When the layer is deployed multiple times, a mechanism should be used that allows to invalidate certain state in all deployed copies, to allow to safely erase authentication state on logout.

\subsubsection{Operations Layer}
\label{operations_layer}
The operations layer offers high level operations possibly involving multiple data objects. It allows to orchestrate multiple data objects, without having to encode complicated choreographies, not really belonging to one object or another, into the data objects resulting in bloated data objects and too tight coupling. As a result of the strict layering and authorisation checking on operation level, the operation layer also has to provide view operations forwarding properties of data objects up to higher layers. However these view operations may bundle data objects often requested at once in one single operation, reducing the total number of requests needed by upper layers and thereby improving performance. Authentication state information also has to be forwarded to the data access authorisation component in the data abstraction layer.

The operations layer should not keep any state and therefore may be replicated, but care should be taken that the transaction component of the data abstraction component is used where required.

\subsubsection{Data Abstraction Layer}
\label{data_abstr_layer}
The data abstraction layer mainly consists of data abstraction objects, providing an object oriented encapsulation to the underlying data assets, like records in a database, certificates and other information. These data abstraction objects also provide methods manipulating the objects, simple operations concerning mainly one object and some constraint checking. This constraint checking at data object level involves for example verifying that there are no valid certificates left when removing a domain, but also checking authorisation on data object level. This authorisation checking is done through the data access authorisation component (which is part of the data abstraction layer). The component uses the information about the authentication state that is forwarded from the authentication and authorisation layer, through the operations layer, to the data abstraction layer, to check whether the particular user may access the object in the requested way.

As duplicated data objects may result in data inconsistencies, even when using transactions (due to higher risk of mistakes in implementation), the data abstraction objects should not be replicated. It would be possible however to partition the data objects to spread the load.

\subsubsection{Back-End/Supporting Services Layer}
\label{backend_layer}
The back-end consists of components that actually store the data represented by the data abstraction objects. That includes data bases, as well as the file system, the audit trail server, and the signing server (for issuing and revoking certificates). The supporting services are components that connect to external services on behalf of the data abstration objects. The data abstraction objects will never directly connect to an external service (for example to verify a domain name), they will always use a component operated by CAcert to act as an intermediary. These may be off-the-shelf components, such as in the case of a mail server which delivers the locally submitted emails to foreign mail servers, or custom components written for this purpose.

\section{Pipes and Filters Architecture}
\label{eval_pipes}

The pipes and filters \gls{AStyle} was evaluated for suitability to fit the requirements. Each request would be interpreted as an object, routed through a system of pipes and filters, eventually being transformed into a response. The problem was that, in order to process the request, quite some filters would add information to the each request, which would then be consumed by later filters. This approach results in many stages of filters and quite some pipes, meaning that data would have to be passed around and possibly be copied quite a lot which makes it inefficient. Also it would result in an unintuitive architecture, which might cause problems in development. Therefore the architecture was not pursued any further.

\section{Service-Oriented Architecture}

The last \gls{AStyle} to be evaluated was the service-oriented architecture. A service oriented architecture is characterised by autonomous components, called services, which provide an interface via an implementation technology independent protocol. These services are combined to provide the functionality of the complete application. By using a protocol that is independent from the implementation, the services can be loosely coupled and implemented in any technology.

\subsection{Choice}
\label{eval_soa_choice}

The service-oriented architecture pattern was chosen because the services can be fully independent, allowing for vertical partitioning of the system, which confines the effects of an attack, and because it is a pattern commonly found in modern business applications.

\subsection{Design}
\label{eval_soa_design}

\begin{figure}[ht]
	\centering
	\includegraphics[width=1.00\textwidth, trim=1.3cm 3.3cm 7.0cm 1.7cm]{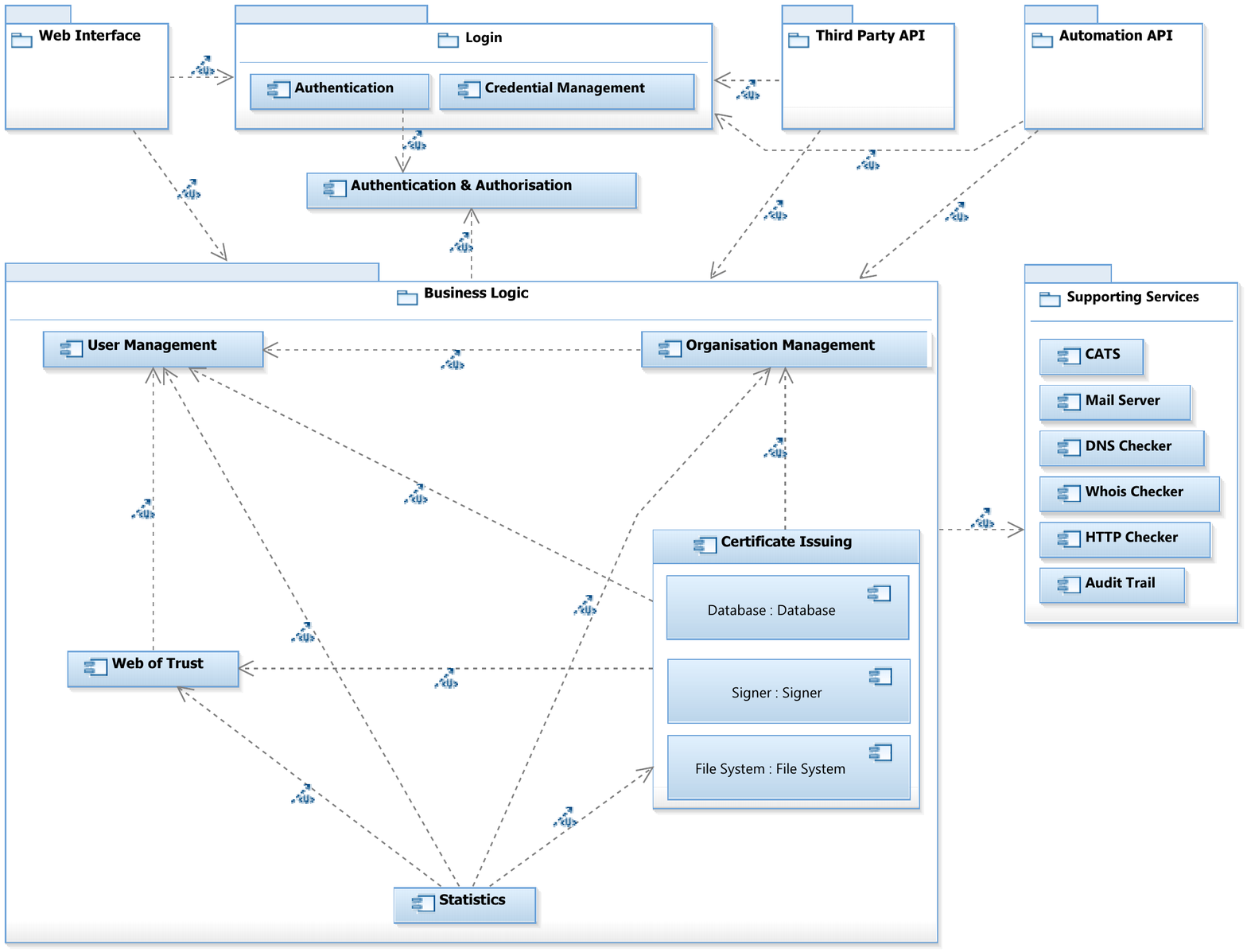}
	\caption{Service-oriented architecture design – larger version in Appendix \ref{soa_design}}
	\label{fig:soa}
\end{figure}

In the design depicted in Figure \ref{fig:soa} the system consists of front-end components, the business logic, an authentication and authorisation service, and supporting services. The general idea is that each service is self-contained, meaning that for example it does not rely on another service for data storage, and cooperates with other services in a choreography. There is no managing component coordinating the cooperation between services. This is different to the layered architecture design, where higher levels mostly rely on lower layers to provide essential operations, such as persisting data, and operations involving multiple data objects are coordinated by the operations layer.

\subsubsection{Front-End}
\label{soa_frontend}
The front-end components are very similar to the components in the front-end layer in the layered architecture design (Section \ref{frontend_layer}), except that interaction context that can't be held by the client is stored in the front-end component itself. This makes it necessary to synchronize the state if the front-end component is deployed on multiple machines.

\subsubsection{Authentication and Authorisation}
\label{soa_auth*}
The authentication and authorisation service is responsible for authentication of clients, managing credentials, and is queried by other services for the authorisation status for each invoked operation. Like in the layers design, authentication tokens are used to keep track of the authentication status. Therefore, they have to be supplied for each invoked operation and are in turn provided to the authentication and authorisation service.

\subsubsection{Business Logic}
\label{soa_business_logic}
The business logic consists of some services providing the core functionality. The services already contain what has been in the back-end in the layers design and connect to supporting services as found in the layers design where needed (see Section \ref{backend_layer}). As already indicated, the collaboration of the services is not managed by another component, but each service is responsible to collaborate with other services of its own accord. For example if the user management service gets a request to remove a domain from the account, it should also invoke the certificate issuing service to revoke all valid certificates issued for that domain. As the services also are responsible for persisting their data, they can't be deployed on multiple machines without some preparation.

\chapter{Evaluation}
\label{ch:Evaluation}

In this chapter the designs presented in Chapter \ref{ch:Alternatives} are evaluated and compared with each other. First, in Section \ref{eval_variation} the variation of the attack tree method used in this work is described. Afterwards, in Sections \ref{eval_layer} and \ref{eval_soa}, the evaluations of both designs are presented and then compared in Section \ref{eval_comparison}. The evaluation shows that there are structural differences in the risk evaluation but no significant differences in the result, even when ignoring a common dominating specification-inherent risk.

\section{Variation of the Evaluation Method}
\label{eval_variation}

The attack tree analysis presented in Section \ref{sec:EvalMethod} is annotated with risk values. Viega and McGraw assign a single risk value to each node that is derived from the estimated time (effort), cost and risk to the attacker \cite{viega2001}. This assessment is slightly modified for this work. Firstly the components making up the risk value are not merged but kept separate, to make it clearer how the risk is derived and make the propagation up in the tree more precise. Secondly time and cost are tightly related in that attacks can usually accelerated if more money is invested and missing capital can be compensated to some point by investing more time, therefore they are combined into a single value called effort. Thirdly a component called gain is added, which represents the motivation for the attacker to carry out the annotated attack – not only in the context of the current goal but in general. This is based on the assumption that an exploit that is useful in many contexts attracts more attackers (\eg it is more valuable to find an exploit in a standard database component or even an encryption protocol than in a custom component used in a single system) and an exploit that results in more capabilities in itself is more valuable for the attacker than one that doesn't and therefore is more likely to be available in any context. The scale of the components ranges from 1 (meaning low effort, risk or gain) to 9 (meaning high effort, risk or gain).

\section{Layered Architecture}
\label{eval_layer}

This sections presents the evaluation of the layered architecture design from Section \ref{eval_layer_design}.

\subsection{Security Evaluation}
\label{eval_layer_security}
An evaluation of the attack risk was conducted according to the attack tree method presented in Section \ref{eval_variation}. The completed attack tree resulting from that evaluation can be found in Appendix \ref{layer_tree}.

The result shows, that the risk to the web of trust, the user data, and the confidentiality of the issued certificates, is dominated by the ability of users to keep their login credentials secret (Section \ref{layers_conf_cred}). This means that the main risk identified lies in the specification/environment of the system and not in the system itself. One solution would be to disallow the use of passwords as means of authentication, but this would violate the requirements and make the system unusable for some users. As some assets can only be accessed by privileged users (\eg the user data may only be modified by support engineers if the account already got assured), one could argue that these users take extra care with their login credentials and the effort for the attacker is higher in these cases, this however doesn't solve the general problem. A mitigation worth exploring, might be requiring two-factor authentication whenever passwords are used (\eg by sending an additional token to the primary email address). But this might have severe consequences if the additional authentication factor fails (for example if the control over the primary email address is lost), because then, even the legitimate user might not be able to log into the account to change the authentication settings.

As the domination by a specification-inherent risk makes comparison between the risk evaluations of different styles less meaningful (they would all show the same result where the domination takes place), it was investigated how the attack tree would change if the dominating risk was assumed to be mitigated in some way. The result is, that now the hijacking of the authentication state (Section \ref{layers_int_wot} Item \ref{layers_hijack_session}) would become the dominant risk, leading to an effort value of 4 and a value for risk to the attacker of 2, for the web of trust, the integrity of login credentials, and other analogous cases. The risk to the confidentiality of the login credentials is dominated by the risk of intercepting the credentials on the login server (Section \ref{layers_conf_cred} Item \ref{layers_creds_intercept_server}), which results in a value for effort and risk to the attacker of 5 and 1 respectively.

\subsection{Overall Evaluation}
\label{eval_layer_summary}
The design was created with scalability in mind, meaning that the upper layers can be replicated onto multiple servers if needed. The components in the back-end may be replicated, depending on their support for this use case. Maintainability and testability should also be good, because of the rather stable interfaces provided by the layers, which confine the effect of code changes \cite{buschmann1996,fielding2000}, and the possibility to emulate lower layers by mock objects. One drawback might be high overhead, caused by the redundant processing of data in each layer \cite{buschmann1996,fielding2000}, which may even be increased if the layers are deployed in tiers, which means additional overhead for serialising and transferring the data between different address spaces or even machines.

\section{Service-Oriented Architecture}
\label{eval_soa}

This sections presents the evaluation of the service-oriented architecture design from Section \ref{eval_soa_design}.

\subsection{Security Evaluation}
\label{eval_soa_security}
Like for the layers design, an evaluation of the attack risk, according to the attack tree method presented in Section \ref{eval_variation}, was conducted. The resulting attack tree can be found in Appendix \ref{soa_tree}.

Unsurprisingly the result shows that, like in the evaluation of the layers design (Section \ref{eval_layer_security}), the risk to the web of trust, the user data, and the confidentiality of the issued certificates, is dominated by the ability of users to protect their login credentials (Section \ref{soa_conf_cred}). If one assumes, for the sake of being able to compare the designs, that the dominating risk of users not being able to protect their credentials would somehow be mitigated, an attacker exploiting the operation logic of a service becomes the dominating risk (Section \ref{soa_int_wot} item \ref{soa_dumb_operation}). Accordingly, the effort the attacker has to make increases to 4 for the web of trust, the integrity of login credentials, and other analogous cases. For the confidentiality of the login credentials the risk of intercepting the credentials on the login server becomes the dominating risk (Section \ref{soa_conf_cred} item \ref{soa_creds_intercept_server}), leading to an effort value of 5.

The finding that the weakness of the design is that an error in a part of the business logic service may compromise the whole service, shows that the expectation when choosing the \gls{AStyle}, that the vertical partitioning might help to confine the effects of an attack, was not entirely met. One cause might be, that the scope of a service is too broad, so that the attack surface (which is all operations offered by that service) is too big and can't compensate for the missing horizontal partitioning. This might be mitigated by narrowing the scope of each service, so that only few operations are dependent on the security of each other, but that is only possible to a certain degree. One limiting factor here, is the data shared by the operations. If two operations working on the same data are split into separate services, then still both services have to access the data somehow and an attack on one of these services gives access to that data shared by both. So this approach only works if the data sets the services operate on still differ in some way, and it has the drawback that invocations that were previously internal to the service become part of the external interface, which especially with service-oriented architectures comes with some overhead, as each request has to be translated into the protocol and then transferred, and also might degrade the cohesion of the component.

One concern resulting from the decoupled structure of the service-oriented architecture, is that an attacker might exploit the connection between the components instead of attacking the components themselves. However this risk is negligible compared to the risks laid out above.

\subsection{Overall Evaluation}
\label{eval_soa_summary}
Due to the principle of every service keeping its own data, it is assumed that deploying a service on multiple machines is generally not possible and requires the service to be built with some sort of synchronisation mechanism. So scalability is not implied by the design and requires some additional work. Maintainability should be good, because the services have well-defined interfaces, which confines coding changes, and the use of an implementation technology independent protocol allows for exchanging a service with another, implemented using a completely different technology. Testability is to a large part dependent on the internal design of a service. Though tests on the service interface level are possible, they are on a rather rough granularity. The communication overhead of the architecture is expected to be quite significant, because each request has to be transformed into the standard protocol and then transferred to the other service, and each operation involves a couple of services (a front-end component, the authentication and authorisation service and at least one business logic service). Due to the decision not to have a component coordinating the cooperation between services (choreography instead of orchestration), there might be situations where there are redundant operations (\eg when deleting a user account, the user management service might request the certificate issuing service to revoke all valid certificates associated with the account and then call its own procedure to delete all domains associated with the account, resulting in another request to the certificate issuing service requesting to revoke all valid certificates issued to that domain, although they already have been revoked). However it is expected that in comparison the overhead resulting from that issue is rather small.

\section{Comparison}
\label{eval_comparison}
Even when ignoring the dominating risk of users not being able to protect their login credentials, the differences in the result of the security evaluation of both designs are marginal (there is a difference in risk to the attacker of one point between the two designs) and are within the accuracy of the method used for evaluation. If anything, the only indicator that can be used is the kind of dominating risk. For the service-oriented architecture design the dominating risk is an attacker being able to exploit the operation logic of a service, which indicates a weakness in the system, while the dominating risk of the layers design lies in the attacker hijacking the authentication state of a user, which is an attack outside the system. When looking at the dominating attack on the service-oriented design, one can see that in the layers design the same attack is limited due to the authorisation checking on the data abstraction layer. So even if an attacker is able to exploit an operation to manipulate data objects in unexpected ways, he still needs to pass the data authorisation checking. On the other hand, when looking at the dominating attack on the layers design, the same attack is also possible in the service-oriented design. So there might be a slight advantage of the layers design from the security point of view, but it really is small.

The maintainability should be about the same for both designs. One aspect where the service-oriented architecture is usually better suited, is if one wants to integrate components implemented using different technologies, as is usually the case with legacy components or components developed in different departments of an organisation (or even not implemented within the organisation at all). However, this use case is probably not needed within CAcert, because there is no part of the legacy application suited for reuse, and CAcert as an organisation is not so big, that it would have different departments doing implementation work independent from each other. Testability is probably a little bit better with the layers design, because of the stable interfaces in the vertical direction, allowing to confine the object under testing. If the layers design is deployed in tiers, the communication overhead is probably higher than in the service-oriented design, because each request path has to pass more boundaries where data is serialised and transferred (at least one time from the topmost layer to the lowest and back again). However the layers design is likely more scalable, as the upper layers can be deployed redundantly while keeping the synchronisation overhead low.

All in all the arguments for the layer design seem stronger in this particular environment and it is therefore recommended for implementation.


\chapter{Conclusion}
\label{ch:Conclusion}

In the evaluated scenario a significant influence of the \gls{AStyle} on the security of the resulting system could not be substantiated. There were some small differences in the resulting risk of certain attacks, but they were too small to distinguish them from possible inaccuracies resulting from the method of evaluation. This does, however, not prove that such an influence does not exist, just that in the evaluated environment there was not enough evidence to prove the existence of such an influence. It was tried to establish rather strict conditions to make a potential positive result stronger and more realistic. The \glspl{AStyle} examined in detail in this work were both chosen for their suitability in security contexts known a priori, if other styles would have been evaluated the evaluation might as well have resulted in a different outcome.

What has been presented in this work, is that a security evaluation can be integrated into a very early stage of the design process. Those security evaluations can show possible attack risks that should be addressed in further design iterations. Integrating security reviews this early in the design process might help to fight the problems that come with the “security as an afterthought” approach.

Because mutliple designs were evaluated in this work, solutions that address certain problems in the earlier designs have been reused in the following design processes. This makes it hard to tell, whether there was an advantage in efficiency in the design process for one \gls{AStyle} or another. Future work could examine this question, for example by designing an experiment measuring the time taken for the design process and security of the resulting system when using different \glspl{AStyle}.

Another interesting question for future investigation would be, whether an influence of the \gls{AStyle} can be shown for more relaxed environments, which still seems likely, or if it can be universally shown to be absent.

Also a different selection of the designs to be evaluated could be promising. For example, evaluating existing systems with the same specification and different architectures for their expected risk for attack, using the method presented in this work, and comparing the result to their real historic security record. Candidates for such an evaluation might be standard components that are regularly scrutinised for their security properties, such as SQL data bases or web servers.

\cleardoublepage
\phantomsection
\addcontentsline{toc}{chapter}{\bibname}

\iflanguage{english}
{\bibliographystyle{IEEEtranSA}}	
{\bibliographystyle{babalpha-fl}}	
												  

\bibliography{papers/thesis}

\cleardoublepage
\printglossaries

\cleardoublepage



\appendix


\chapter{Requirements}
\label{Requirements}

\section{Functional Requirements}

\begin{ind_legal}
	\item Functionality not tied to an user account
		\begin{ind_legal}
			\item Static web pages
				\begin{ind_legal}
					\item Root certificates\oldreq
						\begin{ind_legal}
							\item HTTP download in various formats (PEM, DER, PGP)
							\item Import functionality\\
								Some browsers need extra functionality for easy import of the root certificate (especially Internet Explorer)
							\item Download of the current CRLs\\
								The CRLs might be partitioned in the future to make them smaller in size, so multiple CRLs for the same root have to be supported
						\end{ind_legal}
					
					\item Policies\oldreq\\
						The system should provide a repository of normative policies. This might be moved out of the critical system area in the future. This repository should include the state of the particular policy (work in progress, draft, policy \etc)
					
					\item Statistics\oldreq\\
						The system should provide some statistics about itself (issued certificates, \glspl{Assurer} \etc). The statistics may be regularly pre-generated
				\end{ind_legal}

			\item Contact Support Team\oldreq
				\begin{ind_legal}
					\item List of various places to ask for help\\
						Link to such a list in a non-critical system (\eg the wiki, CMS)
					\item Contact form\\
						This might be moved to a non-critical system (\eg CMS)
						\begin{ind_legal}
							\item Select where to send the message\\
								The user should be able to choose one of various places the message is sent to: Support Engineers for sensitive information and things that can only be done by privileged users, support mailing list for general questions. Other places might be added in the future
							\item Spam protection\newreq
							\item Authorisation token if logged in\newreq\\
								If the user is logged in and sends a message to the Support Engineers include an authorisation token as described in requirement 
\ref{token} and a link to the user account in the message
						\end{ind_legal}
				\end{ind_legal}

			\item Send Emails\altreq\\
				Emails sent by the system should include some easy to configure text in every message (such as donation campaigns) and should be signed with a \gls{x509} certificate that is also used by the support team (this distributes the certificate to many users email clients so they can contact us encrypted).

			\item Internationalisation
				\begin{ind_legal}
					\item Strings and other resources translatable\oldreq\\
						Exceptions possible for interfaces only used by a few persons, such as the Support Engineer interface, where it can be assumed that all persons are capable of understanding English. It has to be able to work with our localisation system (that is Pootle which can handle Gettext, XLIFF, Qt TS, TBX, TMX, Java Properties, Mac OSX strings, PHP arrays and some other formats)
					\item Unicode\newreq\\
						Wherever possible data should be transferred and stored in Unicode aware encoding
					\item Cultural differences should be respected\altreq
				\end{ind_legal}

			\item Accessibility
				\begin{ind_legal}
					\item Unobtrusive JavaScript\oldreq\\
						All features should be accessible without JavaScript or similar techniques enabled. Only convenience features may only work with those technologies turned on.
					\item Accessibility for People with disabilities\newreq\\
						All features should be accessible for people with disabilities (\eg blind people). Where this is not possible alternatives should be available (\eg sound captchas in addition to visual ones)
				\end{ind_legal}

			\item Revocation Status Checking\\
				Third parties need the possibility to obtain the revocation status for certificates issued by CAcert. There are two major techniques for this purpose:
				\begin{ind_legal}
					\item CRLs\label{CRL}\oldreq\\
						CRLs are files that contain a signed list of the serial numbers of revoked certificates. They may be cached by third parties for various reasons within their validity period (for availability or scalability reasons they may also be cached within CAcert but the cache life time should be drastically lower than the validity period of the CRLs). The authoritative CRL should be updated very often (\eg once every 15 minutes) in order to keep the time it takes to propagate the revocation state of a certificate low. In order to keep the size of CRLs low, the CRLs may be partitioned (\eg by month the certificate in question was issued on) but there should be a ``master CRL'' available which contains the revocation status for all or a significant amount (\eg all that haven't expired yet) of issued certificates. This is needed for some configurations (\eg Apache web servers accepting client certificates for authentication).
					
					\item OCSP \label{OCSP}\altreq\\
						OCSP is an interactive protocol that allows a client to query the revocation status from a responder by providing a serial number of the certificate and getting back a signed response valid for a time span specified in that response. If the system does not implement a distributable system of OCSP responders on its own, it has to provide some interface so a separate responder can get the information it needs (at least the revocation status if the serial is provided and some way to successively get a list of all valid serial numbers to be fetched during a very long time span)
				\end{ind_legal}

			\item Advertisements\newreq\\
				It should be possible to include static advertisements hosted at the system in predefined places (plain images or text only, no scripts or interactive objects). Also some statistics interesting for advertisers should be collected (views, clicks \etc; should be modifiable)

			\item User registration\oldreq
				\begin{ind_legal}
					\item Collect user data
						\begin{ind_legal}
							\item One or multiple names\label{register_names}\altreq\\
								Some guidance should be given how these are to be entered.
							\item Date of birth
							\item Acceptance of some agreements\\
								At the moment only the CCA (CAcert Community Agreement), there may be more obligatory agreements in the future
							\item Email address
							\item Credentials for one or more login methods (requirement~\ref{login})\altreq
							\item Credentials for one or more account recovery methods (requirement~\ref{recovery})\altreq\\
								If that is required by the recovery method
							\item Opt-in for announcements/newsletters
						\end{ind_legal}
					
					\item Verify collected data
						\begin{ind_legal}
							\item Verify email address using one or more verification methods (requirement~\ref{email_verification})
							\item Test whether chosen login method works\label{check_login}\\
								\eg for password enter it two times, for challenge/response schemes try one round
						\end{ind_legal}
				\end{ind_legal}
		\end{ind_legal}

	\item Normal user accounts \label{normal}
		\begin{ind_legal}
			\item Login
				\begin{ind_legal}
					\item Several login methods\label{login}\altreq\\
						At least password and client certificate authentication should be offered. Adding of further methods such as one time passwords, hardware tokens \etc should be possible. Also selecting multi-factor authentication should be possible. Which combinations of login methods and the concrete security parameters for each method are acceptable is subject to varying needs and need to be easily adjustable.
					\item Continuous authentication\newreq\\
						When authentication methods where this possible without user interaction are used (\eg client certificate login) the session should be tied to those methods and require reauthentication on every request to hinder session stealing. Also it should be possible to tie the session to a fixed IP (disabling should be possible to allow for the use of anonymisers such as TOR, but maybe only if authentication methods that allow for continuous authentication are used)
				\end{ind_legal}

			\item Account recovery \label{recovery}\altreq\\
				Several methods for account recovery should be offered. It should be possible to deactivate some methods on the own account. Some of them need to be combined to recover the account. Which ones may be/have to be combined should be easily changeable
				\begin{ind_legal}
					\item Password Recovery with \gls{Assurance}\label{PRwA}\newreq\\
						An already in use mechanism currently manually handled by the support engineers. It uses the \gls{Assurance} process to provide an out of band channel with authentication (by passports and ID cards). The procedure is described in\\\url{http://wiki.cacert.org/Support/PasswordRecoverywithAssurance}
					
					\item Control of the primary email address\oldreq\\
						The email verification methods (requirement~\ref{email_verification}) should be used to verify control of the main email address
					
					\item Reset questions\oldreq\\
						A set of questions about stable facts about the user that are nevertheless not known to others and their answers are configured. On recovery the user has to answer some subset of them correct (\eg 3 out of 4). Some generally good questions should be provided (\eg name of first pet), but the user may add her own.
					\item Other Methods\\
						May be added over time.
				\end{ind_legal}

			\item Edit account data\label{edit_account}
				\begin{ind_legal}
					\item Add names\newreq\\
						The same restrictions as for registration (requirement~\ref{register_names}) apply
					\item Remove names\newreq\\
						Names may be marked as deleted at any time, but that also causes certificates in that name to be revoked and the name will only really be deleted when the last certificate expires (even if it already has been revoked). At least one name always has to stay in the account (not necessarily the same name)
					\item Edit date of birth\oldreq\\
						Is only allowed if no name on the account has any \gls{Assurance points}
					\item Add email address\label{email_verification}\altreq\\
						Once an email address is added to the account, the ownership has to be verified by passing at least two checks described below (required by CPS)\todo{it is not clear how this would work on non-primary email addresses, ask Ian about it}. The system should allow for the checks to be repeated if later decided that this is needed (\eg once a year, on every issuing of a certificate if last check too old). If the email address is already assigned to a different account there should be a possibility to transfer it from the old to the new account even if the user has no access to the old account (\eg do the email checks, then send an email to the old accounts primary email address asking to immediately confirm or refuse the transfer within a fixed time span)
						\begin{ind_legal}
							\item Email ping\oldreq\\
								A random string is sent in an email to the address to be verified. This string has to be entered in the system (manually or by clicking a link) and then the user should confirm that he wants to add the email address to the specified account
							\item Email is used in \gls{Assurance} (requirement~\ref{assure})\newreq\\
								That means the \gls{Assurer} has a signed paper form stating that the email address is controlled by the user \(\Rightarrow\) legal instead of technical means
							\item Other Methods\newreq\\
								May be added over time. The system should allow for requiring more methods to verify the email address if policy changes in the future
						\end{ind_legal}
					
					\item Remove email addresses\altreq\\
						Certificates containing that address should be revoked, address should only be marked as deleted until the last certificate expires. At least the primary email address has to remain on the account
					
					\item Select primary email address\oldreq\\
						One email address is the “primary” email address where notifications, announcements and newsletters get sent to
					
					\item Add DNS domain name\label{DNS_verification}\altreq\\
						Once a domain name is added to the account, the ownership has to be verified by passing at least two checks described below (required by CPS). The system should allow for the checks to be repeated if later decided that this is needed (\eg once a year, on every issuing of a certificate if last check too old). If the domain is already assigned to a different account there should be a possibility to transfer it from the old to the new account even if the user has no access to the old account (\eg do the domain checks, then send an email to the old accounts primary email address asking to immediately confirm or refuse the transfer within a fixed time span)
						\begin{ind_legal}
							\item Email ping\oldreq\\
								A random string is sent in an email to a privileged email address on that domain. Privileged addresses are those that are typically only controlled by the administrator or domain owner (those listed in whois, postmaster@domain, \etc). This string has to be entered in the system (manually or by clicking a link) and then the user should confirm that he wants to add the domain name to the specified account
							\item DNS TXT record\newreq\\
								The system generates a random string which needs to be put into a DNS TXT record for the domain. This record is then queried by the system
							\item HTTP\newreq\\
								The system generates a random string which needs to be placed in a file on a web server running on that domain (URL should be specified by the system to avoid Pastebins and other sites allowing to host user content)
							\item Whois\newreq\\
								The system generates a random string which has to be included in the whois database entry for that domain
							\item Statement by 2 \glspl{Assurer}\newreq\\
								Statement by two \glspl{Assurer} about the ownership/control of this domain (only meant as last resort, maybe only under the control of a support engineer?)
							\item Other Methods\newreq\\
								May be added over time. The system should allow for requiring more methods to verify the domain name if policy changes in the future
						\end{ind_legal}
					
					\item Remove domain names\label{rmdomain}\oldreq\\
						Certificates containing that domain name should be revoked, domain should only be marked as deleted until the last certificate expires
					
					\item Modify login method preferences (requirement~\ref{login})\altreq\\
						It should be verified whether the provided credentials work as described in requirement~\ref{check_login}
					
					\item Modify password recovery method preferences (requirement~\ref{recovery})\oldreq\\
						A notification should be sent to the primary email address if critical information is displayed (\eg when viewing common questions for password recovery (“maiden name of the mother” \etc))
					
					\item Modify announcement/newsletter settings\oldreq
				\end{ind_legal}

			\item Web of trust
				\begin{ind_legal}
					\item List \glspl{Assurance}\label{WoTlist}\oldreq\\
						And also the sum of points accumulated on each name. This page should also provide a summary of how to get more points and other requirements to move on. Traditionally there also has been some placement (\eg you are the user with the 178th most \gls{Assurance points}).
					
					\item Link to CATS\oldreq\\
						CATS (CAcert Assurer Testing System) is a system that offers multiple choice test on various subjects, most importantly the Assurer challenge which tests a prospective \gls{Assurer}'s knowledge about the \gls{Assurance} process. The system itself is out of scope of the critical system.
					
					\item Import CATS results\oldreq\\
						The CATS regularly pushes the report about successful tests to the critical system. A passed Assurer challenge is required to become an \gls{Assurer}.
					
					\item List absolved tests\oldreq\\
						The system should provide a list of passed tests so the user can verify if the test done on CATS has been successfully imported
					
					\item Link to the documentation\altreq\\
						The explanation itself about how the WoT works is placed outside the scope of the critical system (\eg CMS, wiki)
					
					\item Links to CAP-form generator\altreq\\
						CAP forms are the forms used in the CAcert \gls{Assurance} Process. The links should go to a pre-filled version (names, date of birth, primary email address encoded in the HTTP request parameters) and a plain version. The CAP form generator itself is placed outside the scope of the critical system.
					
					\item Link to the \gls{Assurer} search\altreq\\
						Link to the system that allows to find an \gls{Assurer} in the same area. The system itself is outside the scope of the critical system (\eg CMS)
				\end{ind_legal}

			\item Trusted Third Party (TTP) programme\newreq\\
				If the user has less than 100 \gls{Assurance points} (precondition may change) there should be a link to the documentation for the TTP programme and a form to request a TTP \gls{Assurance}. If used, a mail should be sent to the TTP Assurer team that includes TTP \glspl{Assurance} already present in the account.

			\item Certificate Management
				\begin{ind_legal}
					\item Requirements common to all certificate types
						\begin{ind_legal}
							\item Certificate signing keys offline\oldreq\\
								The keys for the subroot certificates may only be kept on machines not accessible from the network. Ideally they are additionally contained in a hardware cryptography device (\eg a smartcard)
							\item Only verified information may be included in the certificates\oldreq\\
								Names may only be included if they have been assured with at least 50 \gls{Assurance points} (the concrete preconditions may change). Email addresses and domain names may only be included if they have been verified (requirements \ref{email_verification} and \ref{DNS_verification}). Photo IDs as found in PGP keys and other information may not be signed.
							\item Minimum key size\oldreq\\
								The system should allow to globally specify minimum key sizes for some cryptographic algorithms. No certificate should be issued or renewed if the key doesn't fulfil that requirement
							\item Key restrictions\oldreq\\
								Apart from the key size other restrictions may be applied to the keys. The concrete restrictions should be modifiable.
							\item Validity period\oldreq\\
								Certificates are to be valid for six months for users with less than 50 \gls{Assurance points} and 2 years for others, although the validity periods and point levels may change in the future.
							\item Selection of issuing subroot\oldreq\\
								The decision on which subroot will be used to issue the certificate is subject to change. With the current (as of January 2012) certificates there are only two root certificates: the so-called ``class 1 root'' which may be used for all certificates and the ``class 3 root'' only to be used by users who have more than or exactly 50 \gls{Assurance points} (the user may choose to use the ``class 1 root'' nevertheless). With the new root structure this decision becomes more complex and the exact conditions are documented in the CPS (section §1.4.5)
							\item List of issued certificates\oldreq\\
								The system should provide a list of all certificates issued for the account. To improve orientation it should be possible to hide expired/revoked certificates and at least the following properties should be shown: common name, expiration date, revocation date, serial number, comment entered on creation, link to download
							\item Download of the certificate\oldreq\\
								A possibility to download the certificates should be offered. The download should be restricted to the user itself (and account sitters) for privacy reasons
							\item Renewal of the certificate\oldreq\\
								It should be possible to renew a certificate without going through the whole issuing process again (same properties and public key with another validity period)
						\end{ind_legal}
					
					\item Requirements common to \gls{x509} certificates
						\begin{ind_legal}
							\item Submission via CSR (Certificate Signing Request)\oldreq\\
								The signature on the CSR should be checked and the information contained in it should be taken as default values for the certificate properties in the customisation and confirmation step \todo{possession of private key}
							\item Submission via web browser\oldreq\\
								The keys are interactively generated and the built-in methods to verify the possession of the private key (typically challenge/response based) are used. At least the key size should be selectable.
							\item Customisation and Confirmation of certificate properties\label{customisation}\altreq\\
								The user should confirm (\eg because some properties in the CSR could not be fulfilled -- maybe because of unverified domain names) and be able to edit the properties to be contained in the certificate (up to a certain point, \eg subject alternative names may be edited, key usage may not)
							\item Revocation\altreq\\
								The system should offer a way to revoke certificates (accessible from the certificate list). The revocation date should be set by the user or specify a date set to a date before the validity period. By no means always the current date should be used (if a signature was made before the revocation date it is sometimes considered still valid). Also the user should be able to select the reason why the certificate was revoked (to allow populating the reasons field in the CRL)
						\end{ind_legal}
					
					\item Requirements common to PGP certificates
						\begin{ind_legal}
							\item Submission\oldreq\\
								Because of the nature of PGP, a public key that already contains the properties wanted by the user is uploaded and the system can only check those properties for the concordance with the policies, they can not be edited. PGP keys can contain multiple identities and the system may selectively certify only those that can be verified (the comment field is ignored), this selection should be presented to the user in a confirmation step (including the reasons why the identities can not be certified if there are any), so that it's possible for the user to spot errors and correct them. \todo{Possession of private key}
							\item Revocation\newreq\\
								The user may upload a revocation certificate for each key which can be downloaded at any time by the user. This is basically an escrow service for revocation certificates. To prevent abuse as a file hosting platform and prevent errors, the revocation certificate should be checked for validity for the assigned PGP certificate
							\item Key server upload\newreq\\
								In the future the ability to directly upload the signed key or on revocation a previously provided revocation certificate to a key server may be added
						\end{ind_legal}
					
					\item Certificate types\\
						At least the following types of certificates need to be provided. Further types might be added in the future
						\begin{ind_legal}
							\item \gls{x509} Login certificate\newreq\\
								Does not include any identity information whatsoever and is therefore not useful for third parties but may be used by the member to log in to the CAcert account using certificate authentication (mapping to the account via serial number). This is especially useful as the user can't create other types of client certificates until she is assured at least one time
							\item \gls{x509} Anonymous client certificate\oldreq\\
								Only contains one or more verified email addresses but not a name (``CAcert WoT User'' may be used instead)
							\item \gls{x509} Client certificate\oldreq\\
								Contain a name (which has to match one of the assured names) and any number of verified email addresses
							\item \gls{x509} Server certificate\oldreq\\
								Contain any number of verified domain names and subdomains thereof (no assured name of the member)
							\item PGP certificate\oldreq\\
								Contain any number of Assured name, verified email address pairs
						\end{ind_legal}
				\end{ind_legal}

			\item Account Sitting\newreq\\
				Inexperienced users may give experienced users they trust limited access to their account to help them with certificate issuing, revocation, email/domain verification \etc without giving them their account credentials. Only assured users (\ie users with more than 50 \gls{Assurance points}, though this precondition may change) may act as account sitters. Account sitters may not access or adjust account login or recovery credentials
				\begin{ind_legal}
					\item Explicit enabling\\
						Users should need to explicitly add the account sitter in their account by entering the email address of the account sitter and a notification should be sent to the primary email address of the user including a link/token to confirm the change.
					
					\item List of account sitters\\
						The system should provide a list of persons, with assured name and email address, who have access to the account and provide a possibility to revoke access
					\item Public key submission\\
						The account sitter should be able to specify the properties of the desired certificate but not be able to submit the public key. The user should be provided with a possibility to submit the public key corresponding to the defined properties and get back the issued certificate.
				\end{ind_legal}

			\item Manage third party API permissions\label{manageAPI}\newreq\\
				The user should be able to see all third parties that have access to his account (see requirement~\ref{API}) and what kind of services they are permitted to use. There should be a low-barrier possibility to revoke this access (full or in parts)

			\item Excerpt of all personal data\newreq\\
				To comply with data protection regulations the user should be able to view all personal data CAcert has stored about him
		\end{ind_legal}

	\item \glspl{Assurer}\\
		\gls{Assurer} is a person with 100 or more \gls{Assurance points} (at least one name assured to at least 100 points or at least one name assured by two TTP \glspl{Assurance} and a TTP-TOPUP \gls{Assurance}) and a passed Assurers challenge (these preconditions might change in the future). An \gls{Assurer} should have all the functionality a normal user has (requirement~\ref{normal}) and in addition to that the following:
		\begin{ind_legal}
			\item Assure Someone\label{assure}\\
				When entering the primary email address of a user the \gls{Assurer} should be presented the following data and inputs:
				\begin{ind_legal}
					\item Reminder email\altreq\\
						If the email address entered is not a primary email address of any account in the system, an option should be provided to send an invitation/reminder message to the email address. The language of the message should be selectable and an English version should always be included. If the email address is registered as additional email address of an account but not as primary email address instructions how to solve the problem should be given and the language set in the account should be used. In any case it should instruct the user to contact the \gls{Assurer} once she has created the account or set the primary email address properly or the system should provide a way to notify the \gls{Assurer} automatically once this has happened.
					
					\item Date of Birth\newreq\\
						If the \gls{Assuree} is less than 18 years old a special notice should be shown including a reference to the PoJAM (Policy on Junior Assurers/Members) should be given and the \gls{Assurer} should have to explicitly mark that he did check the requirements given in there (usually parental consent). The exact age from which on this notice is triggered may change.
					
					\item Link to relevant policies/documents\oldreq\\
						That includes Assurance Handbook, Assurance Policy and Practice on Names. More might be added.
					
					\item Names\altreq\\
						All names present in the Account should be shown to the \gls{Assurer} with a possibility to assign each name a certain amount of \gls{Assurance points}. This amount is limited to the maximum amount the \gls{Assurer} may issue which is determined by how many \glspl{Assurance} the \gls{Assurer} has already issued. Current policy is to start with 10 points that may be issued and then for every five \glspl{Assurance} five more points may be issued \todo{maybe switch to one more point per \gls{Assurance} scheme which seems easier to understand?} up to 35 points which is the maximum amount of points that may be issued by a single person. Additionally experience points earned by other means (Assurer Training Events \etc) are taken into account (2 experience points correspond to one \gls{Assurance} given). This policy may however change in the future. If a name has already been assured by the \gls{Assurer} the points may only be increased (\ie if the name was previously assured with 10 points the \gls{Assurer} may increase them up to the maximum amount he can currently issue but not 9 or less points). Directions should be given that this increase is only allowed if actually another \gls{Assurance} did take place, not to correct errors or make use of gained experience points. \glspl{Assurance} to the same user only count once towards the experience level. It should be possible to add functionality to specify in which ID documents the name is contained and show that information to future \glspl{Assurer} (this is to allow detecting faulty \glspl{Assurance} done by other \glspl{Assurer}).
					
					\item Location\oldreq\\
						The location where the \gls{Assurance} took place (free text)
					
					\item Date\oldreq\\
						The date on which the \gls{Assurance} took place. This may be a validated field but the system should respect that in old data from the existing system a free field was used.
					
					\item Agreements the \gls{Assuree} entered into\altreq\\
						There are some Agreements that are captured during the \gls{Assurance} process (currently only the CCA). During the introduction of such Agreements into the process there usually are some \glspl{Assurance} which do not include those Agreements yet (or old \glspl{Assurance} made before the introduction need to be entered). If a particular currently required Agreement is not marked as being entered into by the \gls{Assuree}, the system should ask the \gls{Assurer}, whether this was accidental or if indeed the \gls{Assuree} did not enter into that Agreement. Care should be taken that this question does not lead \glspl{Assurer} to mark the Agreement as entered into while in reality it was not. The set of Agreements the \gls{Assuree} may enter into as well as the subset of those that currently are required may change.
					
					\item Out of band password for requirement~\ref{PRwA}\newreq\\
						Filling this field is optional
					
					\item Statement by the \gls{Assurer}\oldreq\\
						The \gls{Assurer} explicitly needs to mark that he has met the \gls{Assuree} in person and that it was conducted according to the Assurance Policy (for legal reasons) including a link to the relevant policies.
					
					\item Confirmation/Notification Email\oldreq\\
						Once an \gls{Assurance} has been successfully entered a notification email should be send to the \gls{Assurer} and the \gls{Assuree}. The mail to the \gls{Assurer} should contain the points assigned to each name but not the email address of the \gls{Assuree} (to avoid Spammers collecting many addresses if they are able to gain access to an email account of a very active \gls{Assurer}). The email to the \gls{Assuree} should contain the newly gained points as well as the total amount for each assured name and the status of each name (assured/not assured) and if the user has enough points to be an \gls{Assurer} but not passed the \gls{Assurer} challenge yet there should be a pointer to that.
					
					\item Optimisation for multiple \glspl{Assurance}\oldreq\\
						The Assure someone functionality should be optimised for entering multiple \glspl{Assurance} in a row which is often the case for \glspl{Assurer} who are present at open source conventions and similar events. For example date and location should be pre-filled with the data from the previous \gls{Assurance} entered in that session.
				\end{ind_legal}

			\item Link to register as listed \gls{Assurer}\altreq\\
				The registration as listed \gls{Assurer} itself is outside the scope of the critical system (\eg CMS) but there should be a prominent link to get registered.

			\item List of issued \glspl{Assurance}\altreq\\
				Similar to requirement~\ref{WoTlist} the system should provide a list of all \glspl{Assurance} the \gls{Assurer} has entered, how many points the \gls{Assurer} may issue and the position of the \gls{Assurer} in the ranking. If an \gls{Assurance} was entered less than 24 hours ago, a possibility should be provided to send a message requesting the revocation of the \gls{Assurance} to support including the identifier of the \gls{Assurance}, an authorisation token (see requirement~\ref{token}) and a reason why the \gls{Assurance} should be revoked.
			
			\item Request code signing\newreq\\
				An \gls{Assurer} may request to enable code signing certificates for his account. This request triggers an email to support including an authorisation token (requirement~\ref{token}) which then grants or denies the request. In the future this may be automated or enabled by default.
			
			\item \gls{x509} Code signing certificates\oldreq\\
				An \gls{Assurer} who has code signing enabled may in addition to the other certificate types get a code signing certificate. Which is basically a client certificate with a special extension. This may be realised by adding another check box to the customisation and confirmation step (requirement~\ref{customisation}).
		\end{ind_legal}

	\item Organisation Admins\\
		Registered organisations may apply for an organisation account. This is actually not an extended user account but the organisation may name several organisation admins which then get the following additional functionality on their normal user account.
		\begin{ind_legal}
			\item Add another organisation admin\label{addorgadmin}\oldreq\\
				Only users who already are \glspl{Assurer} may be added as organisation admin (this precondition might be adjusted in the future).
				\begin{ind_legal}
					\item Email\\
						To identify the user account to promote to organisation admin
					
					\item Department\\
						The department the new admin is working in \todo{does this actually have any effect or is it just informational?}
					
					\item Comment\\
						An optional comment (\eg why the user has the admin permission) the use of which is free to define by the organisation admins
					
					\item Notifications\newreq\\
						All other admins of the organisation should be notified of the change
				\end{ind_legal}

			\item Remove another organisation admin\label{rmorgadmin}\oldreq

			\item Request addition/removal of a domain name\newreq\\
				On request an email should be sent to the Organisation Assurers who then do the verification and action

			\item Organisation certificates\oldreq\\
				Organisation certificates are \gls{x509} certificates (client, server, code signing) that also contain the organisations details (name, state, country \etc) as specified in requirement~\ref{addorg} Organisation client and code signing certificates are handled a little bit different than in normal user accounts: the email addresses do not need to be verified separately, any email address where the domain part is a domain name of the organisation (or subdomain thereof) is accepted. Also the name is not verified, the organisation admin may just put in any name he sees fit. Organisation client certificates can not be used to log into CAcert accounts

			\item Automation API\newreq\\
				The system should offer an API that allows organisation administrators to automate their issuing process. This API should use special authorisation credentials (\eg tokens or specially marked certificates) that can only be used for accessing the API and not the full account. The system should also offer a way to revoke those credentials and view the last actions performed with those credentials
		\end{ind_legal}

	\item Organisation Assurers\oldreq\\
		Organisations are verified and managed by Organisation Assurers. Organisation Assurers are users who have the ``Organisation Assurer'' role set. Collaboration between Organisation Assurers is done via an external system connected via email (\eg issue tracker or mailing list). Following functionality is available in addition to the features for normal users:
		\begin{ind_legal}
			\item List organisations\\
				List all organisations in the system. There should be means to filter, sort and search this list as it might get quite huge.

			\item Add organisation\label{addorg}\\
				Once the Organisation Assurer has verified the organisation he adds the organisation to the system. The following information has to be included: Name, contact email \todo{is this the first org admin?}, town, state, country (ISO 3166-1 alpha-2 code) and a comment (the comment field has to be big enough to comfortably enter multiple sentences\todo{more structured data?})

			\item Edit organisation\\
				\todo{Only on Arbitration?}

			\item Delete organisation\\
				\todo{Only on Arbitration?}

			\item Manage Organisation Admins\\
				Same as requirements \ref{addorgadmin} and \ref{rmorgadmin}

			\item Add domain name\\
				The ownership will be verified manually by the Organisation Assurer. If the domain name already exists in the system an error should be given.

			\item Remove domain name\\
				Similar to requirement~\ref{rmdomain}
		\end{ind_legal}

	\item Trusted Third Party (TTP) Assurer\\
		The information gathered via TTPs is verified by TTP Assurers. TTP Assurers are users who have the ``TTP Assurer'' role set. The following functionality is available in addition to the features for normal users:
		\begin{ind_legal}
			\item List of TTP \glspl{Assurance}\newreq\\
				On entering the primary email address of a user the TTP Assurer should be able to see a list of names and corresponding TTP \glspl{Assurance} already present on the account.
			
			\item Enter TTP \gls{Assurance}\altreq\\
				On entering the primary email address of a user, the TTP Assurer should be able to add another TTP \gls{Assurance}. If there are already two TTP \glspl{Assurance} present on the account an error message should be given instead. Similar to requirement~\ref{assure} the date of birth and a list of names should be shown. An TTP \gls{Assurance} always gives 35 points to the names that could be verified (concrete amount might change in the future). Similar to requirement~\ref{assure} a list of ID documents which contain the name could be added in the future. Additionally the country, name, location and registration number of the TTP, date of the TTP verification taking place, Agreements the user entered into should be recorded.
		\end{ind_legal}

	\item TTP-TOPUP Assurer\label{TOPUP}\newreq\\
		Via TTP \glspl{Assurance} a user can only gain 70 \gls{Assurance points}. TTP-TOPUP is a programme to educate the user about CAcert's \gls{Assurance} process so he can become a CAcert \gls{Assurer}. This is done by TTP-TOPUP Assurers which are TTP Assurers with the ``TTP-TOPUP Assurer'' role set. The following functionality is available in addition to the features for TTP Assurers:
		\begin{ind_legal}
			\item Enter TTP-TOPUP \gls{Assurance}\\
				In contrast to normal \glspl{Assurance} and TTP \glspl{Assurance} the TTP-TOPUP \gls{Assurance} is not a statement about the identity of the user but the education. Therefore it should apply to the whole account not only single names and does not count towards the verification of names. It only counts towards the \gls{Assurer} capability. On entering the primary email address of a user, the TTP-TOPUP Assurer may mark the account as TTP-TOPUP assured.
		\end{ind_legal}

	\item Arbitrator\newreq\\
		Arbitrators handle difficult cases that are not covered by previously described and authorised procedures. They are the ``judiciary'' of CAcert. Arbitrators may give authorisation for certain actions executed by the Support Engineer. Arbitrators are users who have the ``Arbitrator'' role set. The following functionality is available in addition to the features for normal users:
		\begin{ind_legal}
			\item Issue authorisation tokens\\
				On entering the primary email address of an account and an Arbitration case number, the Arbitrator may select one or more actions to be authorised and get back a token that he can send to support (see requirement~\ref{token}).
			
			\item Manage precedence cases\newreq\\
				The Arbitrator might define a precedence case which has some authorised actions associated with it like a token, but that can be used multiple times provided the executing Support Engineer records the support case number. There also should be a list of existing precedence cases with a possibility to revoke/disable them and view the list of recorded support cases
		\end{ind_legal}

	\item Board Member\\
		Board members are users elected by the community. For the system Board members are users who have the ``Board'' role set. The following functionality is available in addition to the features for normal users:
		\begin{ind_legal}
			\item Issue authorisation tokens to change roles\newreq\\
			On entering the primary email address of an account and a Board motion number, the Board member may select one or more roles assigned or removed and get back a token that he can send to support (see requirement~\ref{token}).

			\item See all privileged users\oldreq\\
				In the privileged roles review (requirement~\ref{privilegereview}) board members always see all privileged users not only those with the same role.
		\end{ind_legal}

	\item Support Engineers\\
		Support Engineers answer help requests by users, execute Arbitration rulings and execute many daily work tasks. They are part of the ``executive'' of CAcert. Support Engineers are users who have the ``Support Engineer'' role set. The following functionality is available in addition to the features for normal users:
		\begin{ind_legal}
			\item View basic details about an account\oldreq\\
				On entering an email address of an account or the account ID at least the following details should be provided but it should be possible to easily add more in the future:
				\begin{ind_legal}
					\item Names\\
						Show all names and the \gls{Assurance points} assigned to them
					
					\item Date of birth
					
					\item Account status\\
						Whether the account has been blocked or blocked from making \glspl{Assurance}
					
					\item Amount of experience points
					
					\item Email addresses\\
						The primary email address should be marked as such
					
					\item Domain names
					
					\item Privileges\\
						Such as assigned roles, if the user is an \gls{Assurer} or not, whether code signing is enabled \etc
					
					\item Passed Tests\\
						Such as Assurers Challenge, Triage Challenge \etc
					
					\item Account activity\\
						A rough guidance about whether the account is still in active used (last login within last month/year \etc) and when the account was initially created
					
					\item Overview over issued certificates\\
						How many certificates have been issued? How many of those have been revoked/expired/are still valid? What is the date when the last certificate expires (even if it has been revoked)?
					
					\item Debugging information\\
						Show technical anomalies in the data records about the user if they are present
				\end{ind_legal}

			\item Privileged functionality\label{token}\newreq\\
				This functionality is only available when a token (a random string that is generated when a privileged action is authorised) is entered or a precedence case is selected and a support case number is entered. It should be easily possible to add more functionality in the future
				\begin{ind_legal}
					\item Read only view of all user details\\
						This should be partitioned by functionality
					
					\item Edit user data\altreq\\
						See requirement \ref{edit_account}. Allow date of birth to be edited even if there are \glspl{Assurance} on some name
					
					\item Revoke \glspl{Assurance}\oldreq\\
						Allow to remove an \gls{Assurance}. The date when the \gls{Assurance} was entered should be displayed
					
					\item Revoke certificates
					
					\item Remove account sitters
					
					\item Remove API permissions
					
					\item Assign or remove roles
				\end{ind_legal}
		\end{ind_legal}

	\item Third party API\label{API}\newreq\\
		The system should offer an API that provides some services to third parties
		\begin{ind_legal}
			\item Only approved parties\\
				The API should only offer the services to registered and approved parties. This approval should also contain the subset of the services the party is generally allowed to use

			\item Only on users permission\\
				If the user has not granted permanent permission (see requirement \ref{manageAPI}) he should be asked if the third party may access his data. This request should contain a list of services the party wants to use with the ability to selectively deny the access and the possibility to choose whether to allow this access once or permanently (until revoked)

			\item Provided services\\
				At least the following services should be provided, others should be easy to add in the future:
				\begin{ind_legal}
					\item Name verification\\
						Given an email address and a name, the system responds whether the data matches. It should only take assured names into account (\ie names that have at least 50 \gls{Assurance points}; concrete precondition might change)
					
					\item Date of birth verification\\
						Given an email address and a date of birth or age the system responds whether the data matches. Only assured accounts are to be taken into account (\ie one name assured to at least 50 \gls{Assurance points}; concrete precondition may change)
					
					\item Age verification\\
						Given an email address, a country and a purpose (\eg old enough to be legally competent, view content not suited for minors, buy alcoholic beverages) the system should respond whether the user is according to the rules in that country allowed for that purpose (only taking age and not other conditions into account). The system should give an error if the age for the country is not known or does not apply. Only assured accounts are to be taken into account (\ie one name assured to at least 50 \gls{Assurance points}; concrete precondition may change)
					
					\item \gls{Assurer} status\\
						Given an email address the system responds whether the user has \gls{Assurer} status
					
					\item \gls{Assurance points} that may be issued\\
						Given an email address the system responds with the amount of \gls{Assurance points} the user may issue (0 if the user is not an \gls{Assurer} yet, see requirement \ref{assure})
				\end{ind_legal}
		\end{ind_legal}

	\item Additional Requirements
		\begin{ind_legal}
			\item Expiration of Assurance and/or Experience points\newreq\\
				The system should allow for changing the point system in such a way, that points expire after some fixed time span or only have an effect if the last \gls{Assurance} is less than a fixed time span in the past (\eg a non-anonymous certificate may only be issued if the last \gls{Assurance} was less than 2 years ago). This might be required by changing external demands.

			\item Review of privileged roles\label{privilegereview}\altreq\\
				All users who have a privileged role should have access to a list containing all other users who also have that same privileged role. Also a regular permission report in the form of an email containing the same information should be sent to those users.
		\end{ind_legal}
\end{ind_legal}

\section{Non-Functional Requirements}

\begin{ind_legal}
	\item Security\\
		As a \gls{CA} security is a major concern. A detailed list of valuable assets and the protection required can be found in the next section
	
	\item Audit trail\\
		All critical operations should be recorded so that in the case of an incident an investigation of the cause can be conducted.
	
	\item Scalability\\
		Once the Audit is complete we expect a substantial rise in users. Therefore it should be easily possible to scale the system by adding further hardware or other easy-to-deploy modifications
	
	\item Maintainability\\
		In the security business often new requirements show up. It should be easy to extend the system to adjust to these changing requirements. Major points where change is already to be expected are mentioned in the functional requirements but other points may also be affected
	
	\item Testability\\
		The system should allow for thorough testing. Testing is employed to ensure the quality of new code and prevent the introduction of regressions
	
	\item Protection of personal data\\
		The protection of the personal data of the users is a major concern to CAcert. Also it is a requirement imposed by legal regulations.
		\begin{ind_legal}
			\item Data economy\\
				Personal data should only be used if needed. And if used it should be used in a way that minimises the impact.
		\end{ind_legal}
	
	\item Usability\\
		In order to facilitate widespread use, the system should be usable by persons of average technical skills.
\end{ind_legal}

\section{Critical Assets}

In this section assets that need special protection are identified along with the kind of protection required and the reason for this protection and the impact if it fails.

\subsection{Web of Trust}

\subsubsection{Integrity}
If the integrity of the web of trust was compromised, an attacker could fake the verification of his identity and therefore create certificates with unverified information in them without having to fear prosecution by backtracking through the web of trust. Changes in the web of trust should be recorded so that after-the-fact investigations if the integrity might have been compromised are possible. If the integrity of the web of trust is compromised and it is not possible to either investigate and recover from the damage or setting the system back to a known safe state this would mean that CAcert would have to rebuild the web of trust from scratch and therefore probably mean the end of existence for the project.

\subsubsection{Confidentiality}
To protect information about relationships between users, which might be considered personal data, the connections in the web of trust should not be public. If the relationship data leaks it might result in bad publicity.

\subsection{Login/Recovery Credentials}

\subsubsection{Integrity}
If an attacker is able to set new credentials he has full access to the account and can issue certificates in that users name and if the user has \gls{Assurer} status he might falsely verify his own account for a fake identity. If multiple accounts get compromised in this way (especially \gls{Assurer} accounts) and it can be determined to which accounts this applies, a whole subgraph might need to be removed from the web of trust which is not desirable but feasible. If it is not possible to determine which accounts have been compromised then essentially the integrity of the web of trust is compromised and the same ways of recovery or failure to do so apply. Changes of login or recovery credentials should be recorded so that an investigation of accounts which might have been compromised is possible.

\subsubsection{Confidentiality}
A user might against all advice use the same password for multiple services so even if CAcert is compromised the credentials stored should not be useful for logging into other services. If an attacker recovers the credentials for an account in a usable way he has full access to the account and might perform any attack described in the previous section. Even if the credentials are recovered in an unusable way it might still result in bad publicity.

\subsection{User Data}

\subsubsection{Integrity}
If an attacker is able to modify the data of his account after he already got Assured or add an email address or domain he does not have control over, he might issue certificates for a fake identity. These fake identity certificates might in turn be used to compromise other systems (\eg secure network connections) and result in a major loss of trust in CAcert certificates. Changes of user data should be recorded so that an investigation of unauthorised changes is possible.

\subsubsection{Confidentiality}
To protect personal data about the user, this information should not be public. \glspl{Assurer} might access some of this data on entering the primary email address in order to assure the user. Failure to enforce confidentiality of private user data might result in bad publicity and many users requesting to delete their data.

\subsection{Issued Certificates}

\subsubsection{Integrity}
While integrity of the certificates themselves stored for the retrieval by the user is not required as they contain inherent integrity checks, some metadata, like which certificate belongs to which user account is critical. If an attacker could associate his certificate with the user account of another user he might be able to use it to login as that user (especially for login-only certificates), effectively resulting in an integrity violation of login credentials. All issuing of certificates should be recorded so that false metadata can be identified.

\subsubsection{Confidentiality}
Certificates should not be public as they might contain personal user data. If the confidentiality is violated it might result in bad publicity for CAcert.

\subsection{Revocation Information}

\subsubsection{Integrity}
If an attacker is able to mark a compromised certificate as not revoked, he still may use this certificate even if the software validating the certificate does revocation checking. Therefore every revocation of a certificate should be recorded so that an investigation and recovery is possible. If such an attack succeeds it might have an impact on trust in CAcert certificates.

\subsubsection{Availability}
If the revocation status of a certificate is not available at the time of verification, the software verifying might react by either warning the user or even marking the certificate invalid or just ignoring this kind of error. Warnings or treating the certificate as invalid might give the impression that the connection is insecure when in reality this is not the case or even worse ignoring the error might lead to putting trust in a certificate that has been compromised. Low availability of revocation information gives the impression that the CAcert service is unreliable, web site owners might complain about inaccessibility for their users and software vendors will ignore revocation status checking errors because of the infeasibility of strict checking. The revocation status providing services should be redundant for these reasons.

\subsection{Root/Subroot Certificates}

\subsubsection{Integrity}
If an attacker is able to replace the root and subroot certificates with his own certificates, users coming to the CAcert website to install the CAcert root certificates might install the attackers certificate. Although there also is the unsolvable problem that the data is changed in transit, this would probably only affect a few connections while replacing the certificates on the website would be persistent and therefore affect more users. The real solution to the problem would be that users verify the fingerprint of the certificate which is distributed out-of-band or receive the certificate in another secure way, but to be realistic one has to assume that this is not done by every user. If an attacker is able to replace the root certificate this might result in bad publicity for CAcert.

\subsection{Certificate Signing Keys}

\subsubsection{Integrity}
If an attacker is able to modify the keys used to sign end user certificates it might pose some interruptions in our service as highly secured backups would have to be restored.

\subsubsection{Confidentiality}
If an attacker is able to retrieve the certificate signing keys or execute some operations on it (\eg signing arbitrary values) this would enable him to sign arbitrary certificates (under the current root setup he might even issue a subroot he might use for himself). A way to recover from this failure might be to revoke the compromised certificate signing keys and generate new ones. All signing operations should be logged to allow investigation and strict input validation is required to hinder signing of dangerous values (\eg a blacklist for critical domain names that may never be contained). Nevertheless if the confidentiality is violated it would result in a major loss of trust in CAcert certificates that would threaten the existence of the project.

\chapter{Layered Design}
\label{layer_design}

\setlength\fboxsep{0pt}
\setlength\fboxrule{0.5pt}
	\includegraphics[angle=90,trim=0.8cm 3.4cm 2.5cm 0.5cm,height=0.8\textheight,width=\textwidth]{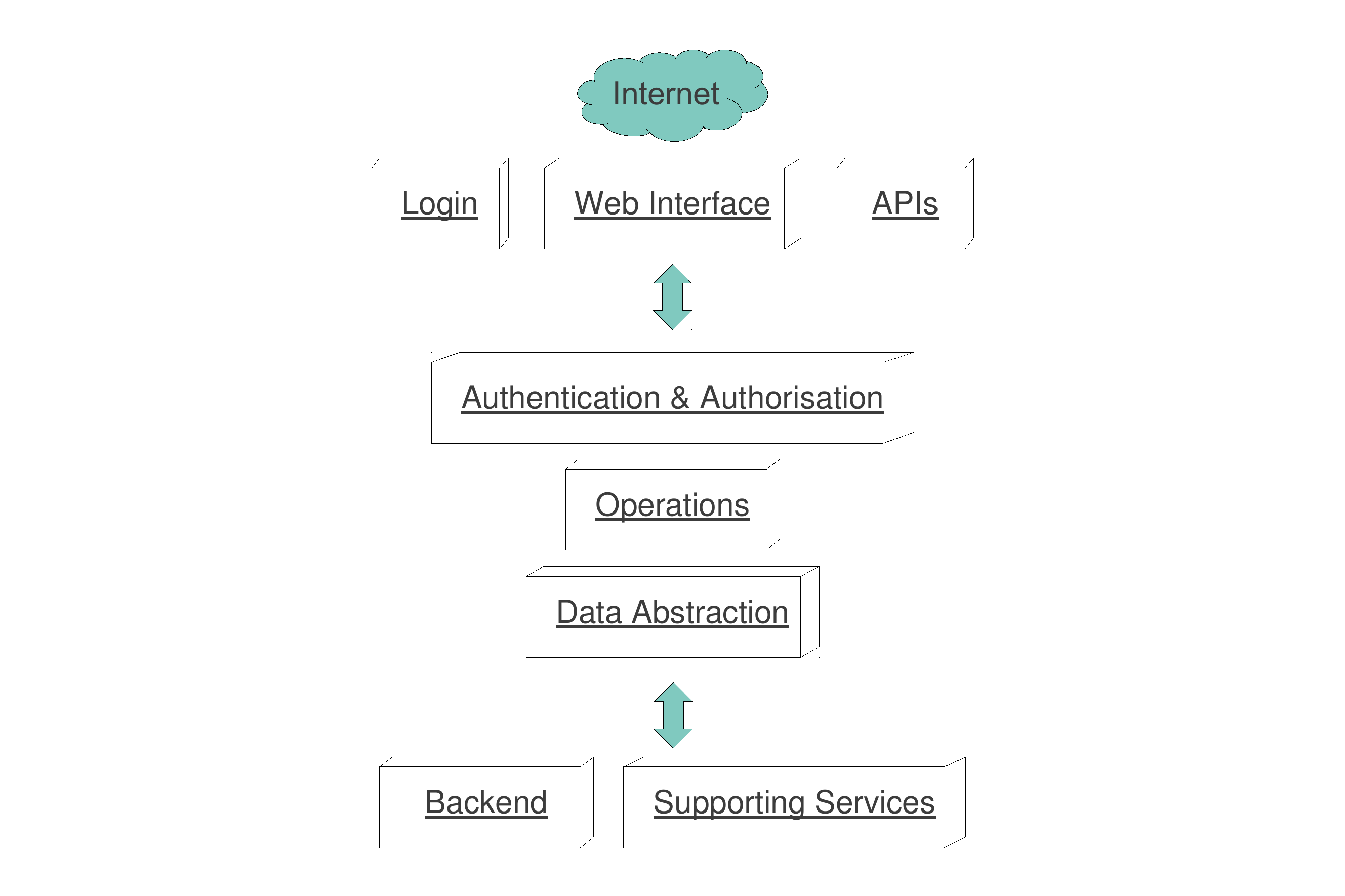}

\chapter{Layered Design Analysis}
\label{layer_tree}

\riskheader

\section{Web of Trust}
\subsection{Unauthorised Modification of the Web of Trust}\risk{2}{1}{8}
\label{layers_int_wot}

\begin{legal}

	\item Direct access to the database\label{layers_db_access}\risk{5}{2}{7}
	\begin{legal}

		\item Recover database credentials AND\risk{5}{1}{3}
		\begin{legal}
			
			\item Recover application credentials\risk{5}{1}{3}
			\begin{legal}
				\item Exploit information leak (\eg via error messages, stack traces and other debugging information)\risk{5}{2}{2}
				\item Exploit memory access violation (buffer overflow attack, printf attack \etc)\risk{6}{1}{8}
				\item File content leak of configuration files (user controlled paths, unintentional check-ins into version control \etc)\risk{5}{1}{3}
				\item Intercept credentials from SQL client connection\risk{8}{2}{5}
				\begin{legal}
					\item Intercept connection on the network AND\risk{4}{2}{4}
					\begin{legal}
						\item Punch hole through firewall\risk{6}{4}{5}
						\item Attack that gains user privileges on hosts on the internal network\risk{4}{2}{5}
					\end{legal}
					\item Attack encryption protocol\risk{8}{1}{9}
				\end{legal}
				\item Set new credentials for application (\eg via SQL injection)\risk{6}{2}{7}
				\item Social Engineering\risk{5}{6}{6}
			\end{legal}
			
			\item Recover administrative credentials\label{layers_admin_cred}\risk{6}{7}{8}
			\begin{legal}
				\item Social Engineering\risk{6}{7}{8}
				\item Attack that gains root privileges on the system running the database (exploit of security vulnerability for the OS or other component running with root privileges, access to hardware \etc)\label{layers_root_access}\risk{7}{2}{8}
			\end{legal}
			
		\end{legal}

		\item Connect to the database\label{layers_conn_host}\risk{4}{2}{3}
		\begin{legal}
			
			\item Punch hole through the firewall and connection exposed to internal hosts\risk{6}{4}{5}
			
			\item Attack that gains user privileges on hosts on the internal network and connection exposed to internal hosts\risk{4}{2}{5}
			
			\item Attack that gains user privileges on the database host\risk{4}{2}{5}
			
		\end{legal}

	\end{legal}

	\item Intercept and modify commands sent to the database\label{layers_int_db_connection}\risk{8}{2}{7}
	\begin{legal}

		\item Database connection exposed to internal hosts AND

		\item Intercept and modify traffic on the internal network AND\risk{4}{2}{4}
		\begin{legal}
			
			\item Punch hole through firewall\risk{6}{4}{5}
			
			\item Attack that gains user privileges on hosts on the internal network\risk{4}{2}{5}
			
			\item Attack that gains root privileges on the database host\risk{7}{2}{8}
			
		\end{legal}

		\item Attack encryption protocol\risk{8}{1}{9}

	\end{legal}

	\item Inject malicious database commands into the commands sent by the data object (SQL injection)\label{layers_SQLinjection}\risk{6}{2}{7}
	\begin{legal}

		\item Bypass input validations at upper layers (\eg by using unusual escape sequences that will be unescaped at lower layers) AND\risk{3}{2}{3}

		\item Execute the injected data in the context of SQL commands (\eg improperly handled variable parts in SQL templates)\risk{6}{1}{4}

	\end{legal}

	\item Modify data objects at the data abstraction layer\risk{2}{1}{7}
	\begin{legal}

		\item Directly modify data objects\label{layers_mod_data_obj}\risk{7}{1}{7}
		\begin{legal}
			
			\item Bypass operations layer (buffer overflow, etc.) AND\risk{6}{1}{4}
			
			\item Pass data access authorisation checking\label{layers_skip_da_auth}\risk{5}{1}{4}
			\begin{legal}
				\item Find a sequence of primitive operations leading to the intended state that are allowed for the attacker\risk{5}{1}{4}
				\item Use authentication state of another user with enough privileges who is currently logged in (\eg memory access violation)\risk{6}{1}{4}
				\item Bypass the checking of authentication state (\eg buffer overflow)\risk{6}{1}{5}
			\end{legal}
			
		\end{legal}

		\item Make an operation the attacker is authorised to execute, modify the data objects in unintended ways\label{layers_dumb_operation}\risk{6}{1}{4}
		\begin{legal}
			
			\item Pass data access authorisation checking (see subtree \ref{layers_skip_da_auth}) AND\risk{5}{1}{4}
			
			\item Exploit operation logic (\eg sloppy condition checking, buffer overflows)\risk{4}{1}{3}
			
		\end{legal}

		\item Execute an operation the attacker is not allowed to execute\label{layers_non_auth_op}\risk{2}{1}{6}
		\begin{legal}
			
			\item Bypass authorisation checking\label{layers_bypass_auth_check}\risk{7}{1}{6}
			\begin{legal}
				\item Bypass authorisation checking on the authentication \& authorisation layer (exploit sloppy error checking, buffer overflows \etc) AND\risk{6}{1}{5}
				\item Pass data access authorisation checking (see subtree \ref{layers_skip_da_auth})\risk{5}{1}{4}
			\end{legal}
			
			\item Cause the authentication \& authorisation layer to call a different operation than the one that has been checked for authorisation\label{layers_different_operation}\risk{7}{1}{6}
			\begin{legal}
				\item Exploit authentication logic (\eg sloppy condition checking, buffer overflows) AND\risk{6}{1}{5}
				\item Pass data access authorisation checking (see subtree \ref{layers_skip_da_auth})\risk{5}{1}{4}
			\end{legal}
			
			\item Use the authentication of another user with the required privileges\risk{2}{1}{6}
			\begin{legal}
				\item Social engineering/cooperation by the user (\eg phishing)\risk{2}{3}{4}
				\item Learn authentication credentials of the user (see section \ref{layers_conf_cred})\risk{2}{1}{6}
				\item Set new authentication credentials for the user (see section \ref{layers_int_cred})\risk{2}{1}{6}
				\item Hijack authentication state of the user\label{layers_hijack_session}\risk{4}{2}{5}
				\begin{legal}
					\item Learn reauthentication credentials (\eg cookies – analogous to \ref{layers_conf_cred}) AND\risk{2}{1}{4}
					\item Pass reauthentication checks, such as IP address restrictions\risk{4}{2}{4}
				\end{legal}
				\item Attack the authentication or reauthentication method (\eg find a way to successfully authenticate without knowing the password)\risk{6}{1}{7}
			\end{legal}
			
		\end{legal}

	\end{legal}

\end{legal}

\subsection{Violation of the Confidentiality of the Web of Trust}\risk{2}{1}{5}
\label{layers_conf_wot}

\begin{legal}

	\item Direct access to database (see section \ref{layers_int_wot} subtree \ref{layers_db_access})\risk{5}{2}{7}

	\item Intercept data received from the database\risk{7}{2}{7}
	\begin{legal}

		\item Database connection exposed to internal hosts AND

		\item Intercept traffic on the internal network AND\risk{4}{2}{4}
		\begin{legal}
			
			\item Punch hole through firewall\risk{6}{4}{5}
			
			\item Attack that gains user privileges on hosts on the internal network\risk{4}{2}{5}
			
			\item Attack that gains root privileges on the database host\risk{7}{2}{8}
			
		\end{legal}

		\item Attack encryption protocol if used\risk{8}{1}{9}

	\end{legal}

	\item Read data from the audit log\risk{6}{2}{6}
	\begin{legal}

		\item Get read access on the internal network\risk{6}{2}{6}
		\begin{legal}
			
			\item Connect to the audit log service (see section \ref{layers_int_wot} subtree \ref{layers_conn_host}) AND\risk{4}{2}{3}
			
			\item Exploit audit log service to read from the log instead of appending to it\risk{6}{1}{3}
			
		\end{legal}

		\item Get read access on the audit trail host\risk{6}{2}{6}
		\begin{legal}
			
			\item Get administrative access on the audit trail host (see section \ref{layers_int_wot} subtree \ref{layers_root_access})\risk{7}{2}{7}
			
			\item Read audit log with user access on the audit trail host\risk{6}{2}{6}
			\begin{legal}
				\item Attack that gains user privileges on the audit trail host AND\risk{4}{2}{3}
				\item Attack that gains read privileges on the audit log file to non-administrative users on the audit trail host (via file content leak or user controlled paths in programmes running as super user \etc)\risk{6}{1}{4}
			\end{legal}
			
		\end{legal}

	\end{legal}

	\item Inject malicious database commands into the commands sent by the data object (SQL injection, see section \ref{layers_int_wot} subtree \ref{layers_SQLinjection})\risk{6}{2}{7}

	\item Access data through the data abstraction layer\risk{2}{1}{6}
	\begin{legal}

		\item Directly access data objects (see section \ref{layers_int_wot} subtree \ref{layers_mod_data_obj})\risk{7}{1}{7}

		\item Make an operation the attacker is authorised to execute, unintentionally disclose data (see section \ref{layers_int_wot} subtree \ref{layers_dumb_operation})\risk{6}{1}{4}

		\item Execute an operation the attacker is not allowed to execute (see section \ref{layers_int_wot} subtree \ref{layers_non_auth_op})\risk{2}{1}{6}

	\end{legal}

\end{legal}

\section{Login Credentials}
\subsection{Unauthorised Modification of the Login Credentials}\risk{2}{1}{6}
\label{layers_int_cred}

\begin{legal}

	\item Direct access to the database (see section \ref{layers_int_wot} subtree \ref{layers_db_access})\risk{5}{2}{7}

	\item Intercept and modify commands sent to the database (see section \ref{layers_int_wot} subtree \ref{layers_int_db_connection})\risk{8}{2}{7}

	\item Inject malicious database commands into the commands sent by the data object (see section \ref{layers_int_wot} subtree \ref{layers_SQLinjection})\risk{6}{2}{7}

	\item Modify credentials at the data abstraction layer\risk{2}{1}{6}
	\begin{legal}

		\item Directly modify data objects (see section \ref{layers_int_wot} subtree \ref{layers_mod_data_obj})\risk{7}{1}{7}

		\item Make an operation the attacker is authorised to execute, modify the data objects in unintended ways (see section \ref{layers_int_wot} subtree \ref{layers_dumb_operation})\risk{6}{1}{4}

		\item Execute an operation the attacker is not allowed to execute (\eg set new login credentials)\risk{2}{1}{6}
		\begin{legal}
			
			\item Bypass authorisation checking (see section \ref{layers_int_wot} subtree \ref{layers_bypass_auth_check})\risk{7}{1}{6}
			
			\item Cause the authentication \& authorisation layer to call a different operation than the one that has been checked for authorisation (see section \ref{layers_int_wot} subtree \ref{layers_different_operation})\risk{7}{1}{6}
			
			\item Use the authentication of another user with the required privileges\risk{2}{1}{6}
			\begin{legal}
				\item Social engineering/cooperation by the user (\eg phishing)\risk{2}{3}{4}
				\item Learn (old) authentication credentials of the user (see section \ref{layers_conf_cred})\risk{2}{1}{6}
				\item Pass account recovery methods (\eg intercept confirmation email sent to the primary email address and learn enough about the user to answer the reset questions)\risk{3}{2}{6}
				\item Hijack authentication state of the user (see section \ref{layers_int_wot} subtree \ref{layers_hijack_session})\risk{4}{2}{5}
				\item Attack the authentication or reauthentication method (\eg find a way to successfully authenticate without knowing the password)\risk{6}{1}{7}
			\end{legal}
			
		\end{legal}

	\end{legal}

\end{legal}

\subsection{Violation of the Confidentiality of the Login Credentials}\risk{2}{1}{7}
\label{layers_conf_cred}

\begin{legal}

	\item Intercept credentials on the machine of the user (trojan, key logger, cross-site-scripting \etc)\risk{3}{2}{6}

	\item Deceive the user to believe a server controlled by the attacker is the server of CAcert\risk{2}{1}{6}
	\begin{legal}

		\item Make the user connect to a server the attacker controls, instead of the real server, and relay the authentication protocol to the real server (phishing, DNS attacks \etc) AND\risk{1}{1}{4}

		\item Bypass server side authentication of the encryption protocol (slightly different spelling in the domain name, attacks against domain name comparison like null characters, use no authentication at all \etc)\risk{2}{1}{6}

	\end{legal}

	\item Intercept credentials on the wire (not possible for challenge/response authentication protocols)\risk{8}{2}{6}
	\begin{legal}

		\item Intercept the connection between the user and the server AND\risk{2}{2}{3}

		\item Break confidentiality of encrypted connection\risk{8}{1}{8}

	\end{legal}

	\item Relay attack on challenge/response protocol\risk{8}{2}{6}
	\begin{legal}

		\item Route connection through a host the attacker controls (\eg by attacking name resolution, routing protocols or social engineering techniques) AND\risk{2}{2}{4}

		\item Break integrity and confidentiality of encrypted connection\risk{8}{1}{9}

	\end{legal}

	\item Intercept credentials on the server\label{layers_creds_intercept_server}\risk{5}{1}{7}
	\begin{legal}

		\item Attack that gains administrative privileges on the login server (similar to section \ref{layers_int_wot} subtree \ref{layers_root_access})\risk{6}{2}{8}

		\item Exploit the login server to store credentials and expose them on successive requests\risk{5}{1}{7}
		\begin{legal}
			
			\item Persist credentials (\eg provoke an error that puts the credential as debug information in a log file) AND\risk{3}{1}{5}
			
			\item Restore persisted credentials (\eg file content leak on a log file)\risk{5}{1}{6}
			
		\end{legal}

		\item Exploit the login server to communicate login credentials to the attacker (\eg via code execution attack)\risk{5}{1}{7}

	\end{legal}

\end{legal}

\section{User Data}

\subsection{Unauthorized Modification of User Data}
\label{layers_int_data}

Analogous to the modification of the web of trust data (section \ref{layers_int_wot}).\risk{2}{1}{8}

\subsection{Unauthorized Access to User Data}
\label{layers_conf_data}

Analogous to exposing data from the web of trust (section \ref{layers_conf_wot}).\risk{2}{1}{5}

\section{Issued Certificates}

\subsection{Modification of Issued Certificates}\risk{5}{1}{7}
\label{layers_int_certs}

\begin{legal}

	\item Direct access to the database with modify access to the certificate information\label{layers_db_mod_cert}\risk{6}{7}{8}
	\begin{legal}

		\item Recover administrative database credentials (see section \ref{layers_int_wot} subtree \ref{layers_admin_cred}) AND\risk{6}{7}{8}

		\item Connect to the database (see section \ref{layers_int_wot} subtree \ref{layers_conn_host})\risk{4}{2}{3}

	\end{legal}

	\item Make the data object report false metadata (\eg the user account the certificate is associated with or whether it may be used for login, by using memory violations, sloppy error checking \etc)\risk{5}{1}{6}

\end{legal}

\subsection{Unauthorized Access to Issued Certificates}
\label{layers_conf_certs}

Analogous to exposing data from the web of trust (section \ref{layers_conf_wot}).\risk{2}{1}{5}

\section{Revocation Information}

\subsection{Unauthorized Modification of Revocation Information}\risk{6}{7}{6}
\label{layers_int_revk}

\begin{legal}

	\item Direct access to the database with modify access to the certificate revocation information\label{layers_db_mod_revk}\risk{6}{7}{8}
	\begin{legal}

		\item Recover administrative database credentials (see section \ref{layers_int_wot} subtree \ref{layers_admin_cred}) AND\risk{6}{7}{8}

		\item Connect to the database (see section \ref{layers_int_wot} subtree \ref{layers_conn_host})\risk{4}{2}{3}

	\end{legal}

	\item Modify the data at the point of distribution into the system of OCSP responders and CRL caches\risk{8}{2}{6}
	\begin{legal}

		\item Forge the signature on the OCSP response or CRL AND\label{layers_forge_revk_sig}\risk{7}{2}{6}
		\begin{legal}
			
			\item Get access to the certificate signing keys (see section \ref{layers_conf_keys})\risk{7}{2}{8}
			
			\item Forge the cryptographic signature\risk{9}{1}{9}
			
		\end{legal}

		\item Modify the data on the wire (man in the middle attack)\risk{8}{2}{7}
		\begin{legal}
			
			\item Route the connection from the OCSP responder or CRL cache to the authoritative revocation server to a host the attacker controls (\eg by attacking name resolution, routing protocols or social engineering techniques) AND\risk{2}{2}{4}
			
			\item Break the authenticity of the encrypted connection\risk{8}{1}{8}
			
		\end{legal}

	\end{legal}

	\item Modify the data in the system of OCSP responders and CRL caches\risk{7}{2}{6}
	\begin{legal}

		\item Forge the signature on the OCSP response or CRL (see subtree \ref{layers_forge_revk_sig}) AND\risk{7}{2}{6}

		\item Get write access to an OCSP responder or CRL cache\risk{4}{2}{6}

	\end{legal}

	\item Modify the data on the wire between the client and the system of OCSP responders and CRL caches\risk{7}{2}{6}
	\begin{legal}

		\item Forge the signature on the OCSP response or CRL (see subtree \ref{layers_forge_revk_sig}) AND\risk{7}{2}{6}

		\item Route the connection from the client to an OCSP responder or CRL cache to a host the attacker controls (\eg by attacking name resolution, routing protocols or social engineering techniques)\risk{2}{2}{6}

	\end{legal}

\end{legal}

\subsection{Prevent Access to Revocation Information}\risk{5}{3}{5}
\label{layers_avail_revk}

\begin{legal}

	\item Route a significant number of connections from clients to OCSP servers or CRL caches to a host the attacker controls or a dead end (\eg by attacking name resolution, routing protocols or social engineering techniques)\risk{5}{5}{6}

	\item Denial of service attack against each of the distributed OCSP responders and CRL caches\risk{6}{3}{5}

	\item Denial of service attack against the point of distribution into the system of OCSP responders and CRL caches (single point of failure but possible to use white list filtering in the router)\risk{5}{3}{3}

	\item Denial of service attack against the signing server (\eg by requesting and revoking many certificates at once)\risk{3}{8}{7}

\end{legal}

\section{Root/Subroot Certificates}

\subsection{Modification of Root/Subroot Certificates}\risk{6}{2}{6}
\label{layers_int_rootcert}

\begin{legal}

	\item Write access to the file system on the front end server\risk{6}{2}{8}
	\begin{legal}

		\item Attack that gains root privileges on the front end server (similar to section \ref{layers_int_wot} subtree \ref{layers_root_access})\risk{6}{2}{8}

		\item Exploit some component or service running with elevated privileges on the front end server to overwrite files\risk{6}{2}{8}

	\end{legal}

\end{legal}

\section{Certificate Signing Keys}

\subsection{Modification of the Certificate Signing Keys}\risk{8}{2}{6}
\label{layers_int_keys}

\begin{legal}

	\item Exploit some component or service running with elevated privileges on the signing server to overwrite the certificate signing keys AND\risk{8}{2}{6}

	\item Access the serial connection (see section \ref{layers_conf_keys} subtree \ref{layers_serial_access})\risk{6}{2}{8}

\end{legal}

\subsection{Violation of the Confidentiality of the Certificate Signing Keys}\risk{7}{2}{8}
\label{layers_conf_keys}

\begin{legal}

	\item Exploit the signing server via the serial access protocol\risk{7}{2}{8}
	\begin{legal}

		\item Exploit the signing server to expose information about the key material (\eg by causing part of the signing key material to be included in a signed certificate via memory violation) AND\risk{7}{2}{8}

		\item Access the serial connection\label{layers_serial_access}\risk{6}{2}{8}
		\begin{legal}
			
			\item Directly access the serial connection (\eg by getting root access on the host running the data abstraction layer, see section \ref{layers_int_wot} subtree \ref{layers_root_access})\risk{7}{2}{8}
			
			\item Route malicious data through the normal signing process (similar to the SQL injection, see section \ref{layers_int_wot} subtree \ref{layers_SQLinjection})\risk{6}{2}{7}
			
		\end{legal}

	\end{legal}

	\item Gather information about the signing key material by observing side channels\risk{7}{5}{7}
	\begin{legal}
		
		\item Discover a side channel (timings, power consumption, cache access patterns \etc) AND\risk{5}{2}{4}
		
		\item Access the side channel (\eg by analysing response times or measuring power consumption)\risk{7}{5}{4}
		
	\end{legal}

\end{legal}

\chapter{Service-Oriented Design}
\label{soa_design}

\setlength\fboxsep{0pt}
\setlength\fboxrule{0.5pt}
	\includegraphics[angle=90,trim=1.3cm 3.3cm 7.0cm 1.7cm,height=0.8\textheight,width=\textwidth]{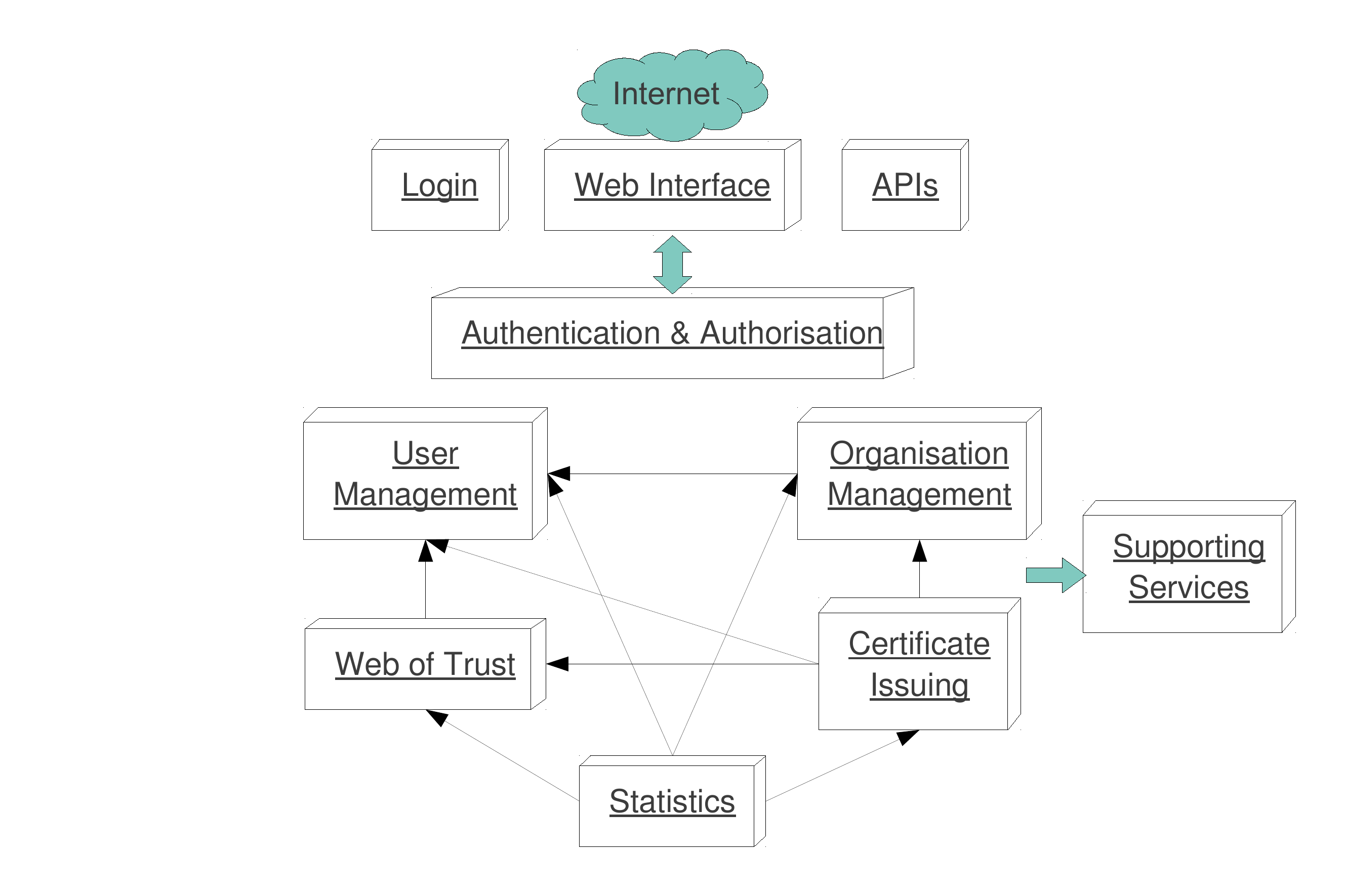}

\chapter{Service-Oriented Design Analysis}
\label{soa_tree}

\riskheader

\section{Web of Trust}
\subsection{Unauthorised Modification of the Web of Trust}\risk{2}{1}{8}
\label{soa_int_wot}

\begin{legal}

	\item Direct access to the database on the web of trust host\label{soa_db_access}\risk{5}{2}{7}
	\begin{legal}

		\item Recover database credentials AND\risk{5}{1}{3}
		\begin{legal}
			
			\item Recover application credentials\risk{5}{1}{3}
			\begin{legal}
				\item Exploit information leak (\eg via error messages, stack traces and other debugging information)\risk{5}{2}{2}
				\item Exploit memory access violation (buffer overflow attack, printf attack \etc)\risk{6}{1}{8}
				\item File content leak of configuration files (user controlled paths, unintentional check-ins into version control \etc)\risk{5}{1}{3}
				\item Set new credentials for application (\eg via SQL injection)\risk{6}{2}{7}
				\item Social Engineering\risk{5}{6}{6}
			\end{legal}
			
			\item Recover administrative credentials\label{soa_admin_cred}\risk{6}{7}{8}
			\begin{legal}
				\item Social Engineering\risk{6}{7}{8}
				\item Attack that gains root privileges on the system running the database (exploit of security vulnerability for the OS or other component running with root privileges, access to hardware \etc)\label{soa_root_access}\risk{7}{2}{8}
			\end{legal}
			
		\end{legal}

		\item Connect to the database\label{soa_conn_host}\risk{4}{2}{3}
		\begin{legal}
			
			\item Attack that gains user privileges on the web of trust host\risk{4}{2}{5}
			
		\end{legal}

	\end{legal}

	\item Inject malicious database commands into the commands sent by the web of trust component (SQL injection)\label{soa_SQLinjection}\risk{6}{2}{7}
	\begin{legal}

		\item Bypass input validations (\eg by using unusual escape sequences that will be unescaped at components closer to the web of trust component) AND\risk{3}{2}{3}

		\item Execute the injected data in the context of SQL commands (\eg improperly handled variable parts in SQL templates)\risk{6}{1}{4}

	\end{legal}

	\item Modify data in the web of trust component\risk{2}{1}{7}
	\begin{legal}

		\item Make a web of trust operation the attacker is authorised to execute, modify the data in unintended ways\label{soa_dumb_operation}\risk{4}{1}{4}
		\begin{legal}
			
			\item Exploit operation logic (\eg sloppy condition checking, buffer overflows)\risk{4}{1}{3}
			
		\end{legal}

		\item Execute an operation the attacker is not allowed to execute\label{soa_non_auth_op}\risk{2}{1}{6}
		\begin{legal}
			
			\item Bypass authorisation checking in the web of trust component\label{soa_bypass_auth_check}\risk{6}{1}{5}
			
			\item Make the authorisation component report that the attacker is allowed to execute the operation\label{soa_report_as_authorised}\risk{7}{3}{7}
			\begin{legal}
				\item Directly exploit the authorisation or authentication component\risk{7}{3}{7}
				\item Man-in-the-middle attack on the connection to the authorisation component\label{soa_mitm_component}\risk{8}{2}{5}
				\begin{legal}
					\item Intercept connection to the authorisation component AND\risk{4}{2}{5}
					\begin{legal}
						\item Punch hole through firewall\risk{6}{4}{5}
						\item Attack that gains user privileges on hosts on the internal network\risk{4}{2}{5}
					\end{legal}
					\item Attack authentication/encryption protocol between the two components\risk{8}{1}{9}
				\end{legal}
			\end{legal}
			
			\item Use the authentication of another user with the required privileges\risk{2}{1}{6}
			\begin{legal}
				\item Social engineering/cooperation by the user (\eg phishing)\risk{2}{3}{4}
				\item Learn authentication credentials of the user (see section \ref{soa_conf_cred})\risk{2}{1}{6}
				\item Set new authentication credentials for the user (see section \ref{soa_int_cred})\risk{2}{1}{6}
				\item Hijack authentication state of the user\label{soa_hijack_session}\risk{4}{2}{5}
				\begin{legal}
					\item Learn reauthentication credentials (\eg cookies – analogous to \ref{soa_conf_cred}) AND\risk{2}{1}{4}
					\item Pass reauthentication checks, such as IP address restrictions\risk{4}{2}{4}
				\end{legal}
				\item Attack the authentication or reauthentication method (\eg find a way to successfully authenticate without knowing the password)\risk{6}{1}{7}
			\end{legal}
			
		\end{legal}

	\end{legal}

	\item Modify the data in transit to the front end or another component (\eg the certificate issuing component, see subtree \ref{soa_mitm_component})\risk{8}{2}{5}

\end{legal}

\subsection{Violation of the Confidentiality of the Web of Trust}\risk{2}{1}{5}
\label{soa_conf_wot}

\begin{legal}

	\item Direct access to database on the web of trust host (see section \ref{soa_int_wot} subtree \ref{soa_db_access})\risk{5}{2}{7}

	\item Read data from the audit log\risk{6}{2}{6}
	\begin{legal}

		\item Get read access on the internal network\risk{6}{2}{6}
		\begin{legal}
			
			\item Connect to the audit log service AND\risk{4}{2}{3}
			\begin{legal}
				\item Punch hole through firewall\risk{6}{4}{5}
				\item Attack that gains user privileges on hosts on the internal network\risk{4}{2}{5}
			\end{legal}
			
			\item Exploit audit log service to read from the log instead of appending to it\risk{6}{1}{3}
			
		\end{legal}

		\item Get read access on the audit trail host\risk{6}{2}{6}
		\begin{legal}
			
			\item Get administrative access on the audit trail host (see section \ref{soa_int_wot} subtree \ref{soa_root_access})\risk{7}{2}{8}
			
			\item Read audit log with user access on the audit trail host\risk{6}{2}{6}
			\begin{legal}
				\item Attack that gains user privileges on the audit trail host AND\risk{4}{2}{3}
				\item Attack that gains read privileges on the audit log file to non-administrative users on the audit trail host (via file content leak or user controlled paths in programmes running as super user \etc)\risk{6}{1}{4}
			\end{legal}
			
		\end{legal}

	\end{legal}

	\item Inject malicious database commands into the commands sent by the web of trust component (SQL injection, see section \ref{soa_int_wot} subtree \ref{soa_SQLinjection})\risk{6}{2}{7}

	\item Access data in the web of trust component\risk{2}{1}{6}
	\begin{legal}

		\item Make a web of trust operation the attacker is authorised to execute, unintentionally disclose data (see section \ref{soa_int_wot} subtree \ref{soa_dumb_operation})\risk{4}{1}{4}

		\item Execute an operation the attacker is not allowed to execute (see section \ref{soa_int_wot} subtree \ref{soa_non_auth_op})\risk{2}{1}{6}

	\end{legal}

\end{legal}

\section{Login Credentials}
\subsection{Unauthorised Modification of the Login Credentials}\risk{2}{1}{6}
\label{soa_int_cred}

\begin{legal}

	\item Direct access to the database on the authentication host (see section \ref{soa_int_wot} subtree \ref{soa_db_access})\risk{5}{2}{7}

	\item Inject malicious database commands into the commands sent by the authentication component (SQL injection, see section \ref{soa_int_wot} subtree \ref{soa_SQLinjection})\risk{6}{2}{7}

	\item Modify credentials in the authentication component\risk{2}{1}{6}
	\begin{legal}

		\item Make an operation of the authentication component the attacker is authorised to execute, modify the data in unintended ways (see section \ref{soa_int_wot} subtree \ref{soa_dumb_operation})\risk{4}{1}{4}

		\item Execute an operation the attacker is not allowed to execute (\eg set new login credentials)\risk{2}{1}{6}
		\begin{legal}
			
			\item Bypass authorisation checking in the authentication component\risk{6}{1}{5}
			
			\item Make the authorisation component report that the attacker is authorised to execute the operation (see section \ref{soa_int_wot} subtree \ref{soa_report_as_authorised})\risk{7}{3}{7}
			
			\item Use the authentication of another user with the required privileges\risk{2}{1}{6}
			\begin{legal}
				\item Social engineering/cooperation by the user (\eg phishing)\risk{2}{3}{4}
				\item Learn (old) authentication credentials of the user (see section \ref{soa_conf_cred})\risk{2}{1}{6}
				\item Pass account recovery methods (\eg intercept confirmation email sent to the primary email address and learn enough about the user to answer the reset questions)\risk{3}{2}{6}
				\item Hijack authentication state of the user (see section \ref{soa_int_wot} subtree \ref{soa_hijack_session})\risk{4}{2}{5}
				\item Attack the authentication or reauthentication method (\eg find a way to successfully authenticate without knowing the password)\risk{6}{1}{7}
			\end{legal}
			
		\end{legal}

	\end{legal}

\end{legal}

\subsection{Violation of the Confidentiality of the Login Credentials}\risk{2}{1}{7}
\label{soa_conf_cred}

\begin{legal}

	\item Intercept credentials on the machine of the user (trojan, key logger, cross-site-scripting \etc)\risk{3}{2}{6}

	\item Deceive the user to believe a server controlled by the attacker is the server of CAcert\risk{2}{1}{6}
	\begin{legal}

		\item Make the user connect to a server the attacker controls, instead of the real server, and relay the authentication protocol to the real server (phishing, DNS attacks \etc) AND\risk{1}{1}{4}

		\item Bypass server side authentication of the encryption protocol (slightly different spelling in the domain name, attacks against domain name comparison like null characters, use no authentication at all \etc)\risk{2}{1}{6}

	\end{legal}

	\item Intercept credentials on the wire (not possible for challenge/response authentication protocols)\risk{8}{2}{6}
	\begin{legal}

		\item Intercept the connection between the user and the server AND\risk{2}{2}{3}

		\item Break confidentiality of encrypted connection\risk{8}{1}{8}

	\end{legal}

	\item Relay attack on challenge/response protocol\risk{8}{2}{6}
	\begin{legal}

		\item Route connection through a host the attacker controls (\eg by attacking name resolution, routing protocols or social engineering techniques)\risk{2}{2}{4}

		\item Break integrity and confidentiality of encrypted connection\risk{8}{1}{9}

	\end{legal}

	\item Intercept credentials on the server\label{soa_creds_intercept_server}\risk{5}{1}{7}
	\begin{legal}

		\item Attack that gains administrative privileges on the login server (similar to section \ref{soa_int_wot} subtree \ref{soa_root_access})\risk{6}{2}{8}

		\item Exploit the login server to store credentials and expose them on successive requests\risk{5}{1}{7}
		\begin{legal}
			
			\item Persist credentials (\eg provoke an error that puts the credential as debug information in a log file) AND\risk{3}{1}{5}
			
			\item Restore persisted credentials (\eg file content leak on a log file)\risk{5}{1}{6}
			
		\end{legal}

		\item Exploit the login server to communicate login credentials to the attacker (\eg via code execution attack)\risk{5}{1}{7}

	\end{legal}

\end{legal}

\section{User Data}

\subsection{Unauthorized Modification of User Data}
\label{soa_int_data}

Analogous to the modification of the web of trust data (section \ref{soa_int_wot}).\risk{2}{1}{8}

\subsection{Unauthorized Access to User Data}
\label{soa_conf_data}

Analogous to exposing data from the web of trust (section \ref{soa_conf_wot}).\risk{2}{1}{5}

\section{Issued Certificates}

\subsection{Modification of Issued Certificates}\risk{5}{1}{7}
\label{soa_int_certs}

\begin{legal}

	\item Direct access to the database of the certificate issuing component with modify access to the certificate information\label{soa_db_mod_cert}\risk{6}{7}{8}
	\begin{legal}

		\item Recover administrative database credentials (see section \ref{soa_int_wot} subtree \ref{soa_admin_cred}) AND\risk{6}{7}{8}

		\item Connect to the database (see section \ref{soa_int_wot} subtree \ref{soa_conn_host})\risk{4}{2}{3}

	\end{legal}

	\item Make the certificate issuing component report false meta data (\eg the user account the certificate is associated with or whether it may be used for login)\risk{5}{1}{6}
	\begin{legal}

		\item Exploit the certificate issuing component to report the false meta data (memory violation, sloppy error checking \etc)\risk{5}{1}{6}

		\item Modify the data in transit (\eg to the authentication component, see section \ref{soa_int_wot} subtree \ref{soa_mitm_component})\risk{8}{2}{5}

	\end{legal}

\end{legal}

\subsection{Unauthorized Access to Issued Certificates}
\label{soa_conf_certs}

Analogous to exposing data from the web of trust (section \ref{soa_conf_wot}).\risk{2}{1}{5}

\section{Revocation Information}

\subsection{Unauthorized Modification of Revocation Information}\risk{6}{7}{6}
\label{soa_int_revk}

\begin{legal}

	\item Direct access to the database of the certificate issuing component with modify access to the certificate revocation information\label{soa_db_mod_revk}\risk{6}{7}{8}
	\begin{legal}

		\item Recover administrative database credentials (see section \ref{soa_int_wot} subtree \ref{soa_admin_cred}) AND\risk{6}{7}{8}

		\item Connect to the database (see section \ref{soa_int_wot} subtree \ref{soa_conn_host})\risk{4}{2}{3}

	\end{legal}

	\item Modify the data at the point of distribution into the system of OCSP responders and CRL caches\risk{8}{2}{6}
	\begin{legal}

		\item Forge the signature on the OCSP response or CRL AND\label{soa_forge_revk_sig}\risk{7}{2}{6}
		\begin{legal}
			
			\item Get access to the certificate signing keys (see section \ref{soa_conf_keys})\risk{7}{2}{8}
			
			\item Forge the cryptographic signature\risk{9}{1}{9}
			
		\end{legal}

		\item Modify the data on the wire (man in the middle attack)\risk{8}{2}{7}
		\begin{legal}
			
			\item Route the connection from the OCSP responder or CRL cache to the authoritative revocation server to a host the attacker controls (\eg by attacking name resolution, routing protocols or social engineering techniques) AND\risk{2}{2}{4}
			
			\item Break the authenticity of the encrypted connection\risk{8}{1}{8}
			
		\end{legal}

	\end{legal}

	\item Modify the data in the system of OCSP responders and CRL caches\risk{7}{2}{6}
	\begin{legal}

		\item Forge the signature on the OCSP response or CRL (see subtree \ref{soa_forge_revk_sig}) AND\risk{7}{2}{6}

		\item Get write access to an OCSP responder or CRL cache\risk{4}{2}{6}

	\end{legal}

	\item Modify the data on the wire between the client and the system of OCSP responders and CRL caches\risk{7}{2}{6}
	\begin{legal}

		\item Forge the signature on the OCSP response or CRL (see subtree \ref{soa_forge_revk_sig}) AND\risk{7}{2}{6}

		\item Route the connection from the client to an OCSP responder or CRL cache to a host the attacker controls (\eg by attacking name resolution, routing protocols or social engineering techniques)\risk{2}{2}{6}

	\end{legal}

\end{legal}

\subsection{Prevent Access to Revocation Information}\risk{5}{3}{5}
\label{soa_avail_revk}

\begin{legal}

	\item Route a significant number of connections from clients to OCSP servers or CRL caches to a host the attacker controls or a dead end (\eg by attacking name resolution, routing protocols or social engineering techniques)\risk{5}{5}{6}

	\item Denial of service attack against each of the distributed OCSP responders and CRL caches\risk{6}{3}{5}

	\item Denial of service attack against the point of distribution into the system of OCSP responders and CRL caches (single point of failure but possible to use white list filtering in the router)\risk{5}{3}{3}

	\item Denial of service attack against the signing server or certificate issuing component (\eg by requesting and revoking many certificates at once)\risk{3}{8}{7}

\end{legal}

\section{Root/Subroot Certificates}

\subsection{Modification of Root/Subroot Certificates}\risk{6}{2}{6}
\label{soa_int_rootcert}

\begin{legal}

	\item Write access to the file system on the front end server\risk{6}{2}{8}
	\begin{legal}

		\item Attack that gains root privileges on the front end server (similar to section \ref{soa_int_wot} subtree \ref{soa_root_access})\risk{6}{2}{8}

		\item Exploit some component or service running with elevated privileges on the front end server to overwrite files\risk{6}{2}{8}

	\end{legal}

\end{legal}

\section{Certificate Signing Keys}

\subsection{Modification of the Certificate Signing Keys}\risk{8}{2}{6}
\label{soa_int_keys}

\begin{legal}

	\item Exploit some component or service running with elevated privileges on the signing server to overwrite the certificate signing keys AND\risk{8}{2}{6}

	\item Access the serial connection (see section \ref{soa_conf_keys} subtree \ref{soa_serial_access})\risk{6}{2}{8}

\end{legal}

\subsection{Violation of the Confidentiality of the Certificate Signing Keys}\risk{7}{2}{8}
\label{soa_conf_keys}

\begin{legal}

	\item Exploit the signing server via the serial access protocol\risk{7}{2}{8}
	\begin{legal}

		\item Exploit the signing server to expose information about the key material (\eg by causing part of the signing key material to be included in a signed certificate via memory violation) AND\risk{7}{2}{8}

		\item Access the serial connection\label{soa_serial_access}\risk{6}{2}{8}
		\begin{legal}
			
			\item Directly access the serial connection (\eg by getting root access on the host running the certificate issuing component, see section \ref{soa_int_wot} subtree \ref{soa_root_access})\risk{7}{2}{8}
			
			\item Route malicious data through the normal signing process (similar to the SQL injection, see section \ref{soa_int_wot} subtree \ref{soa_SQLinjection})\risk{6}{2}{7}
			
		\end{legal}

	\end{legal}

	\item Gather information about the signing key material by observing side channels\risk{7}{5}{7}
	\begin{legal}
		
		\item Discover a side channel (timings, power consumption, cache access patterns \etc) AND\risk{5}{2}{4}
		
		\item Access the side channel (\eg by analysing response times or measuring power consumption)\risk{7}{5}{4}
		
	\end{legal}

\end{legal}

\end{document}